\documentclass[12pt]{iopart}

\usepackage{iopams}  
\usepackage{hyperref}
\usepackage{graphicx}

\usepackage{color,soul}
\newcommand{\IR}{\mathbb{R}}

\begin{document}

\title[Spatially Homogeneous Universes with Late-Time Anisotropy]{Spatially Homogeneous Universes with Late-Time Anisotropy}

\author{Andrei Constantin$^1$\footnote{Author to whom any correspondence should be addressed.}, Thomas R.~Harvey$^1$,
Sebastian von Hausegger$^2$, and Andre Lukas$^1$}
\address{$^1$ Rudolf Peierls Centre for Theoretical Physics, University of Oxford, Parks Road, Oxford, UK}
\address{$^2$ Astrophysics, University of Oxford, Denys Wilkinson Building, Keble Road, Oxford, UK}

\ead{andrei.constantin@physics.ox.ac.uk, thomas.harvey@physics.ox.ac.uk, sebastian.vonhausegger@physics.ox.ac.uk, and andre.lukas@physics.ox.ac.uk}

\vspace{10pt}
\begin{indented}
\item[]August 2023
\end{indented}

\begin{abstract}
The Cosmological Principle asserts that on sufficiently large scales the universe is homogeneous and isotropic on spatial slices. To deviate from this principle requires a departure from the FLRW ansatz. In this paper we analyse the cosmological evolution of two spatially homogeneous but anisotropic universes, namely the spatially closed Kantowski-Sachs universe and the open axisymmetric Bianchi type III universe. These models are characterised by two scale factors and we study their evolution in universes with radiation, matter and a cosmological constant. In all cases, the two scale factors evolve differently and this anisotropy leads to a lensing effect in the propagation of light. We derive explicit formulae for computing redshifts, angular diameter distances and luminosity distances and discuss the predictions of these models in relation to  observations for type Ia supernovae and the CMB. We comment on the possibility of explaining the observed luminosity distance plot for Type Ia supernovae within the context of cosmologies featuring late-time anisotropy and a vanishing cosmological constant.

\end{abstract}

\vspace{2pc}

\submitto{\CQG}
\maketitle

\section{Introduction}
The picture painted by the standard $\Lambda$CDM model about the universe is built on the assumptions of spatial homogeneity and isotropy on very large scales and throughout the entire cosmological evolution, collectively referred to as the cosmological principle~\cite{deSitter:1934,1937RSPSA.158..324M}. These assumptions are validated a posteriori by the success of the model in explaining the accelerated expansion of the universe, the structure of the CMB, the abundances of the light elements and the large-scale structure. On the other hand, the $\Lambda$CDM model, essentially a six-parameter fit to current observations~\cite{Planck:2018vyg}, relies on ingredients that still lack solid theoretical understanding such as dark matter~\cite{Bertone:2004pz} and dark energy~\cite{Copeland:2006wr,Peebles:2002gy,Sarkar:2022qgb}.  Moreover, tensions between early-universe and late-universe measurements, such as the Hubble and the $S_8$ tensions~\cite{Perivolaropoulos:2021jda, DiValentino:2021izs} as well as the presence of large angle anomalies in the CMB data~\cite{Copi:2010na} motivate exploring models beyond the scope of the standard $\Lambda$CDM model~\cite{Abdalla:2022yfr}.

Of particular relevance to this work are a number of observations that directly challenge the two pillar assumptions of homogeneity and isotropy. These include the discovery of structures~\cite{Lopez:2022kbz} significantly larger than the scale of homogeneity found in $\Lambda$CDM simulations~\cite{Yadav:2010cc}, as well as various findings of large-scale anisotropies, such as direction-dependent scaling relations of X-ray galaxy clusters~\cite{Migkas:2020fza,Migkas:2021zdo}, excess dipolar modulations of number counts of radio galaxies~\cite{Blake:2002gx,Rubart:2013tx} and mid-IR quasars~\cite{Secrest:2020has,Secrest:2022uvx,Dam:2022wwh}, and anisotropies in expansion as well as acceleration from type Ia supernovae~\cite{Schwarz:2007wf,Javanmardi:2015sfa,Colin:2019opb}, to name a few (see also~\cite{Aluri:2022hzs}). Despite being faced with a plethora of challenges~\cite{Abdalla:2022yfr}, the $\Lambda$CDM model retains its unquestionable advantage of computational simplicity; the relatively simple picture provided by the model is to a large degree consistent with observations, while any departure from homogeneity and isotropy necessarily complicates the analysis, making it difficult to infer and test observational consequences of non-FLRW models. Nevertheless, pursuing such an analysis seems worthwhile in the attempt to resolve the current tensions between astrophysical data and $\Lambda$CDM-predictions. Since FLRW models provide a good first approximation to current observations, it makes sense to consider only those models that include an FLRW limit, either parametrically or for a range of their cosmological evolution. 

In the present study we concentrate exclusively on homogeneous, non-isotropic models, relaxing the cosmological principle but retaining the Copernican principle~\cite{Misner1968, Collins1973}. For a review of inhomogeneous cosmological models generalising the FLRW universe see~\cite{Bolejko_2011}. Homogeneous cosmologies fall into two classes. The first class consists of the Bianchi models, for which the isometry group admits a 3-dimensional simply transitive subgroup. There are 9 Bianchi-type models, two of which form uncountable classes. Out of these, types I, V, VII$_h$, and IX can be studied as homogeneous linear perturbations on top of an isotropic universe~\cite{Pontzen:2010eg}. For instance, the potential use of Bianchi models in explaining the large-angle anomalies in the CMB has been known for a while, for instance in~\cite{Jaffe_2005, Pontzen:2007ii} it has been shown that the quadrupole-octupole alignment and the low quadrupole moment are simultaneously reduced by a significant amount in the Bianchi type VII$_h$ model. However, CMB data turned out to be inconsistent with this model~\cite{Planck:2015gmu}. CMB constraints on Bianchi I models~\cite{Russell_2014, Akarsu:2019pwn, Akarsu:2021max}, Bianchi V models and Bianchi IX models~\cite{2016PhRvL.117m1302S, Saadeh:2016bmp} have also been obtained.

The second class of homogeneous non-isotropic cosmologies corresponds to the Kantowski-Sachs (KS) models~\cite{Kompanyeets:1964, Kantowski:1966te}, for which the isometry group is neither simply transitive nor does it admit a simply transitive subgroup~\cite{Collins:1977fg, wainwrightellis1997,Ellis:1998ct,ellis_maartens_maccallum_2012}. Concretely, these can be found among those space-times where the topology of the space-like hypersurface is of the form $X_1 \times X_2 = \mathbb R\times S^2$ or other topologies derived from this by identifications of points under translations in the $X_1$-direction or identifications of antipodal points in~$S^2$ or a combination of these. A prominent example is $S^1\times S^2$. In particular, we will consider cases where the metric on $X_1$ is maximally symmetric. If $X_2$ is replaced by a flat  2-dimensional maximally symmetric space, one recovers the axisymmetric Bianchi type I model. If $X_2$ is replaced by a closed 2-dimensional maximally symmetric space, the result is an axisymmetric Bianchi type III model. In this paper we will focus on the spatially closed and the open cases, from here onwards referred to simply as the closed and open cases.  Written in a synchronous frame, the metric for these models takes the form:
\begin{equation}
\label{eq:metric}
ds^2 = -  dt^2 + b(t)^2 r_1^2 d\chi^2
 + a(t)^2 r_2^2 (d\theta^2 + {\rm si}^2(\theta)d\phi^2)~,
\end{equation}
where $t$ is the time coordinate, $\chi\in[0,\infty)$ is the radial coordinate and $\theta\in[0,\pi]$ and $\phi\in[0,2\pi]$ are the spherical angular coordinates. We use a system of units where $c= 1$. The constants $r_1$ and $r_2$ are introduced such that $\chi, \theta$ and $\phi$ are dimensionless. In the flat case, ${\rm si}(\theta) = 1$; in the open case ${\rm si}(\theta) = \sinh(\theta)$, while ${\rm si}(\theta) = \sin(\theta)$ corresponds to the closed KS universe. 

The KS universe may arise in string cosmology~\cite{Barrow:1996gx, Adamek:2010sg} by considering models where all space dimensions are initially compact and small and three dimensions become macroscopic through spontaneous de-compactification. The closed KS universe and the axisymmetric Bianchi type~III universe (which, by slight abuse of terminology, we will refer to as the open KS universe) do not contain FLRW as a special case, as the two-dimensional curvature is always present and breaks maximal three-dimensional symmetry. However, an approximate FLRW model is obtained if the two-dimensional curvature is small compared to the other terms in the Einstein equations (see below), in which case the two scale factors $a$ and $b$ evolve at approximately the same rate.

Anisotropy is typically studied as a feature of the early universe. In this study, however, we explore the possibility that anisotropy is a recent phenomenon by studying the open and closed KS universes.  The propagation of light in an increasingly anisotropic background necessarily influences all cosmological observables, in particular the CMB. However, since the CMB temperature anisotropies are of approximate $10^{-5}$ amplitude, the present-day deviation from isotropy must still be very small. Our aim here is to
provide certain bounds on the open and closed KS universes filled with an isotropic perfect fluid moving orthogonally to the hypersurfaces of homogeneity. Lastly, it is important to note that the maximal symmetry of $X_2$ does not allow for a tilted cosmology, in which the fluid vector would not be normal to the hypersurfaces of homogeneity.  In this context we emphasise that this work does not attempt to perform an exhaustive coverage of all possible homogeneous models with late-time anisotropy.  Instead, we explore to a deeper extent the present subset of models in terms of their theoretical and observational implications.

The paper is organized as follows. In Section~\ref{sec:EE} we begin by writing the Einstein equations for the closed and the open KS universes and then proceed to  solving them in the three cases of matter domination, matter and cosmological constant mixture, and radiation domination. In Section~\ref{sec:light_prop} we discuss the propagation of light in such backgrounds, deriving explicit formulae for the red-shift, the angular diameter distance and the luminosity distance as measured by an observer in an open/closed KS universe filled with matter or matter and cosmological constant. The observational consequences of the lensing effect caused by the anisotropic geometry are explored in Section~\ref{sec:observables}. In particular we show that the accelerated expansion of the Universe can be seen as an apparent phenomenon caused by the lensing properties of the background. However, the required amount of anisotropy turns out to be in conflict with the CMB anisotropy. We conclude in Section~\ref{sec:conclusions}.

\section{Einstein equations}\label{sec:EE}
We consider a KS universe with topology $\IR{\times}X_1{\times}S^2$ or $\IR{\times}X_1{\times}H_2$  where $X_1$ is $S_1$ or~$\mathbb R$. The metric $g_{\mu\nu}$ corresponds to the choice ${\rm si}(\theta)=\sin(\theta)$ in the line element of~\eref{eq:metric}, for the closed KS universe and ${\rm si}(\theta)=\sinh(\theta)$ for the open KS universe. The model is characterized by two scale factors, $a(t)$ and $b(t)$ and two Hubble rates
\begin{equation}
H_a = \frac{1}{a(t)}\frac{da(t)}{dt}\;,\quad H_b = \frac{1}{b(t)}\frac{db(t)}{dt}~.
\end{equation}

In the following, it will be convenient to write 
\begin{equation}
	a(t)=e^{A(t)}\;,\quad b(t)=e^{B(t)}\; .
\end{equation}
and to introduce a new time coordinate $\tau$ related to $t$ by $dt = e^{N(\tau)}d\tau$. Suitable choices of the lapse function $N(\tau)$ will lead to certain simplifications below. With these redefinitions, the metric becomes
\begin{equation}
	ds^2 = - e^{2N( \tau)} d\tau^2 + e^{2B(\tau)} r_1^2 d\chi^2 + e^{2 A(\tau)}r_2^2 (d\theta^2 + {\rm si}^2(\theta)d\phi^2)~.
\end{equation}

For simplicity we set $r_1=r_2$ (that is, by absorbing their ratio into the definition of $\chi$). Assuming that the universe is filled with a co-moving isotropic perfect fluid~\footnote{A fluid velocity in the $\chi$ direction is, a priori, possible but given our choice of coordinates this velocity is forced to zero by the $t$-$\chi$ component of the Einstein equation.}
with energy density $\rho$, pressure $p$, and equation of state $p=\omega \rho$, the conservation of the stress-energy tensor implies the continuity equation
\begin{equation}
	\rho' + (2A'+B')(\rho+p)=0~,
\end{equation}
where the prime indicates a derivative with respect to $\tau$. 
This integrates to
\begin{equation}
	\rho =\hat\rho e^{-(1+\omega)(2A+B)}~=\frac{\hat\rho}{a^{2(1+w)}b^{1+w}}~,
\end{equation}
where $\hat\rho$ is a positive integration constant. 

The Einstein field equations take the form 
\begin{eqnarray}\label{eqn:EFE}
	\eqalign{
	{A'}^2+2A'B'=\kappa\,\rho\, e^{2N}-\frac{k}{r_2^2}e^{2N-2A}\\
	2A''+3{A'}^2-2A'N'=-\kappa\, p\,e^{2N}-\frac{k}{r_2^2}e^{2N-2A}\\
	A''{+}B''{+}{A'}^2{+}{B'}^2{+}A'B'{-}A'N'{-}B'N'{=}-\kappa\, p\, e^{2N}
	}
\end{eqnarray}
where $\kappa = 8\pi G$, $k=1$ for the closed KS universe and $k=-1$ for the open KS universe. 

If we introduce an `averaged' Hubble rate $\tilde{H}$ by
\begin{equation}\label{Htilde}
 	\tilde{H}^2=\frac{1}{3}(H_a^2+2H_aH_b)\; ,
\end{equation}
then the first Einstein equation can be written in the usual form
\begin{equation}
	3\tilde{H}^2:=H_a^2+2H_aH_b=\kappa\,\rho-\frac{k}{r_2^2a^2}\; .
\end{equation}
or, alternatively, as
\begin{equation}
	\Omega-1=\frac{k}{3r_2^2a^2\tilde{H}^2}\;,\quad \Omega=\frac{\rho}{\rho_c}\;,\quad \rho_c=\frac{3\tilde{H}^2}{\kappa}\; .
\end{equation}
However, note that the `critical density' $\rho_c$ is defined relative to this averaged Hubble rate $\tilde{H}$ and, therefore, has a different meaning than in the standard case. In particular, the quantity $\tilde{H}^2$ can become negative. As such, the usual relationship between $k$ and~$\Omega$ only follows under the assumption that $\tilde{H}^2$ is positive and this is certainly the case whenever $X_2$ and $X_1$ are both expanding or both contracting. 

Another obvious approach towards `averaging' the two Hubble parameters is to define the averaged scale factor
\begin{equation}\label{Hbar}
	\bar{a}:=(a^2b)^{1/3}\quad\Rightarrow\quad \bar{H}:=e^{-N}\frac{\bar{a}'}{\bar{a}}=\frac{1}{3}(2H_a+H_b)\; .
\end{equation}
Clearly, if $a=b$, then the two averaged Hubble rates are the same, $\bar{H}=\tilde{H}$, but otherwise we can write
\begin{equation}\label{rho_ghost}
	3\tilde{H}^2=3\bar{H}^2-\frac{1}{3}(H_a-H_b)^2=3\bar{H}^2-\sigma^2
\end{equation}
where $\sigma^2$ is simply the shear scalar~\cite{Pontzen:2010eg}. The Friedman equation can then be re-written in terms of $\bar{H}$ as
\begin{equation}
	3\bar{H}^2=\kappa\rho+\sigma^2-\frac{k}{r_2^2a^2}\; ,
\end{equation}
which shows that interpreting such a universe in the FLRW paradigm with some averaged Hubble rate  may lead to a Friedman equation with an additional `energy density' $\sigma^2\kappa^{-1}$. As the explicit solutions below show this has an expansion of the form
\begin{equation}\label{rho_ghost1}
	\sigma^2/\kappa=\frac{\bar{\rho}_2}{\bar{a}^2}+\frac{\bar{\rho}_1}{\bar{a}}+\bar{\rho}_0+\cdots\;
\end{equation}
where the $\bar{\rho}_i$ are constants. The leading term proportional to $\bar{a}^{-2}$ contributes to the curvature term, as  expected from a model whose primary source of anisotropy is curvature, but there are further terms with lower inverse powers of $\bar{a}$, including an effective cosmological constant term. It is, therefore, not far-fetched to think that the anisotropy of the model, when interpreted in the context of an FLRW model, can mimic a cosmological constant.

In the following, we present special solutions to the Einstein Equations~\ref{eqn:EFE} for the various cases considered here, before we proceed to compute cosmological observables such as luminosity and angular diameter distances.  At the present stage, the interested reader is referred to Ref.~\cite{Stephani:2003tm} where an exhaustive list of general solutions is collected; for the presented cases see also~\cite{Bradley:2011rt}.

\subsection{Solutions with matter}
In this case $p=0$, $\rho=\hat{\rho} e^{-2A-B}$ and it is useful to choose the gauge $N=A$. The Einstein equations~\eref{eqn:EFE} then become
\begin{eqnarray}\label{k1matter}
	\eqalign{
	{A'}^2+2A'B'=\kappa\,\hat{\rho}\, e^{-B}-\frac{k}{r_2^2}\\
	2A''+{A'}^2=-\frac{k}{r_2^2}\\
	A''+B''+{B'}^2=0
	}
\end{eqnarray}
The choice $N=A$ implies that the second equation can be directly integrated to obtain~$A(\tau)$. Inserting the result into the first equation then leads to $B(\tau)$. 
\vspace{4pt}

\begin{figure}[htb]
	\centering
	\includegraphics[width=0.48\textwidth]{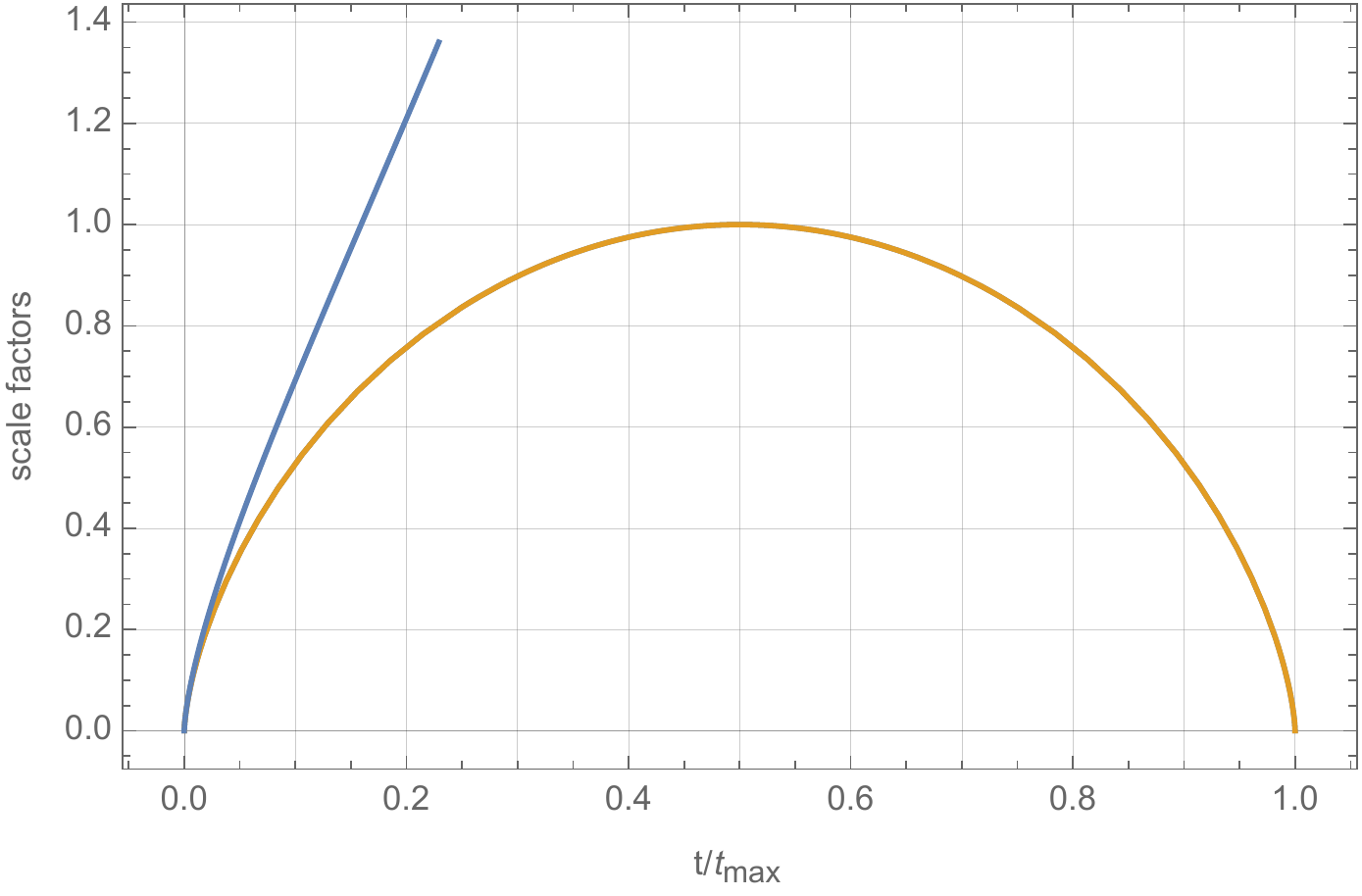}
	\caption{The evolution of the scale factors $a(t/t_{\rm max})/a_{\rm max}$ (orange curve, corresponding to the $S^2$ part of the metric) and $3b(t/t_{\rm max})/b_{\rm max}$ (blue curve, corresponding to the $X_1$ part of the metric) in matter dominated closed KS universe. By choosing $a_{\rm max}=b_{\rm max}/3$, the scale factors $a(t)$ and $b(t)$ have the same asymptotic form near $t=0$, $a(t)\simeq b(t)\sim t^{2/3}$.}
	\label{fig:MatterPlotClosed1}
\end{figure}

{\bfseries The closed KS universe.} We discuss first the case $k=1$ case and write the solutions in terms of the variable $\eta = \tau/r_2$:
\begin{eqnarray}\label{k1mattersol}
	\eqalign{
	a(\eta)=e^{A(\eta)}=a_{\rm max}\sin^2\left(\frac{\eta}{2}\right)\;,\\
	b(\eta)=e^{B(\eta)}=b_{\rm max}\left[1-\frac{\eta-\eta_\ast}{2}\cot\left(\frac{\eta}{2}\right)\right]\;,
	}
\end{eqnarray}
where $b_{\rm max}:=\kappa\,\hat{\rho}\, r_2^2$, while $a_{\rm max}$, $\eta_\ast$ are integration constants and the trivial time-shift integration constant has been set to zero (but can be re-instated by replacing $\eta\rightarrow \eta-\eta_0$). The relation between time $\tau$ (or $\eta$) and co-moving time is obtained by integrating $dt=a(\tau) d\tau$ which leads to
\begin{equation}
	\frac{t}{t_{\rm max}}=\frac{1}{2\pi}(\eta-\sin(\eta))\;,\qquad t_{\rm max}:=\pi r_2a_{\rm max}\; .
\end{equation}
The time parameter $\eta$ runs in the range $\eta\in [0,2 \pi]$, which corresponds to the co-moving time range $t\in [0,t_{\rm max}]$. The $S^2$ scale factor vanishes at either end of the interval, $a(0)=a(2\pi)=0$. It
increases in the first half, $\eta\in [0,\pi]$, to a maximum of $a(\pi)=a_{\rm max}$,  and then collapses in the second half $\eta\in [\pi,2\pi]$, as expected for a closed universe.
Depending on the behaviour of the scale factor $b$ near $\eta=0$, we distinguish between two types of solutions that have an approximate FRLW limit: 
\vspace{4pt}

{\bfseries Type I solutions}: $\eta_\ast=0$. It is convenient to set $a_{\rm max}=b_{\rm max}/3$ in order to match the asymptotic behaviour of the two scale factors near $t=0$. In this scenario the universe starts in the isotropic regime $a(t)\simeq b(t)\sim t^{2/3}$ and the degree of anisotropy increases monotonically in time (see the discussion below). The evolution of the scale factors $a(t)$ and $b(t)$ is shown in \fref{fig:MatterPlotClosed1}. 
\vspace{4pt}

\begin{figure}[htb]
	\centering
	\includegraphics[width=0.48\textwidth]{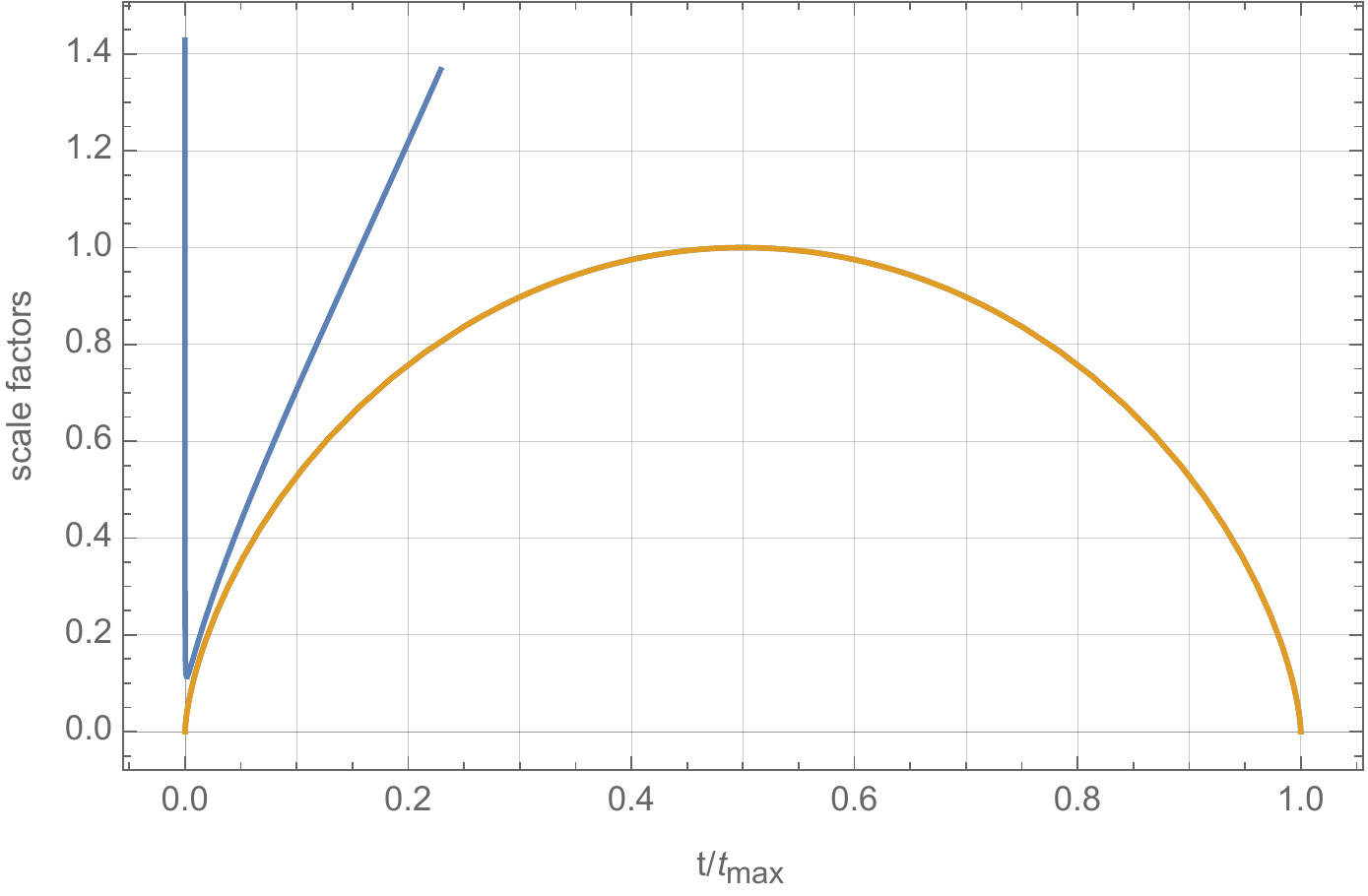} 	\hspace{10pt}
	\includegraphics[width=0.48\textwidth]{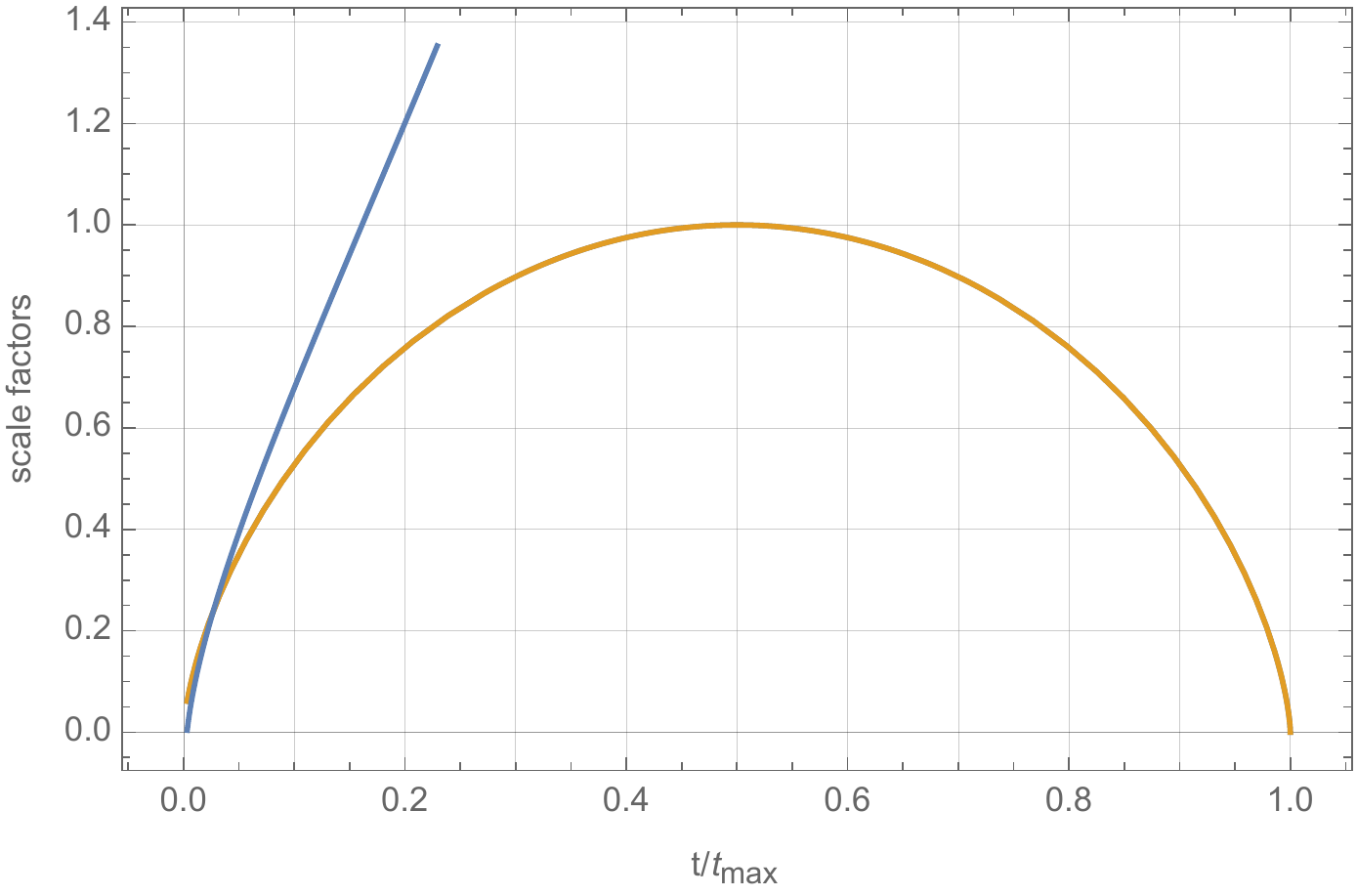}
	\caption{The evolution of the scale factors in matter dominated closed KS universe when $\eta_*\neq0$. On the left $\eta_*=0.01$, while on the right $\eta_\ast=-0.01$. Both plots show $a(t/t_{\rm max})/a_{\rm max}$ (orange curve, corresponding to the $S^2$ part of the metric) and $3b(t/t_{\rm max})/b_{\rm max}$ (blue curve, corresponding to the $X_1$ part of the metric). The universe is nearly isotropic in the range $1\gg\eta\gg\eta_*$.}
	\label{fig:MatterPlotClosed2}
\end{figure}

{\bfseries Type II solutions}: $\eta_\ast\neq 0$. When $\eta\gg |\eta_\ast|$, the evolution of $b$ is similar to the case $\eta_\ast=0$. On the other hand, for small values of $\eta$ the evolution depends on the sign of~$\eta_\ast$.   Thus, for $\eta_\ast> 0$, the $X_1$ scale factor $b$ diverges at $\eta=0$ due to the singularity of the cotangent. 
Initially $b$ contracts from infinity, while $a$ expands from zero. For $\eta_\ast\ll1$, there is an intermediate regime, corresponding to $\eta_\ast\ll \eta\ll 1$, in which the two scale factors evolve approximately at the same rate. At later times the scale factors diverge again, due to the effect of the two-dimensional curvature term in Einstein's equations. 
An illustration of this scenario is shown in the left hand side plot in \fref{fig:MatterPlotClosed2} for $\eta_*=0.01$. Note, the `bouncing' behaviour of the scale factor $b$ does not constitute a theoretical problem, for example in view of the weak energy condition. For multiple scale factors, this behaviour is, in fact common. For a realistic cosmology, the matter-dominated solution from figure~\ref{fig:MatterPlotClosed2} should be patched onto a radiation dominated one at early times. It remains to be seen whether this can lead to a uniformly expanding~$b$. 

For $\eta_\ast<0$, $b$ goes to zero at a finite positive value of $\eta$, which we denote by $\eta_{\rm min}$. As such, the range for $\eta$ should be restricted to $[\eta_{\rm min}, 2\pi]$. In this scenario the universe starts in a two-dimensional phase. An illustration of this scenario is shown in the right hand side plot in  \fref{fig:MatterPlotClosed2} for $\eta_*=-0.01$.

It will be useful to expand the solutions \eref{k1mattersol} around $\eta=0$. The expansion for the scale factor $b(\eta)$ has a regular part which depends on $\eta$ alone and a singular part proportional to $\eta_\ast/\eta$:
\begin{eqnarray}\label{solk1}
\eqalign{
\fl	a(\eta)=\frac{a_{\rm max}}{4}\eta^2\left(1-\frac{\eta^2}{12}+\frac{\eta^4}{360}\right)+{\cal O}(\eta^8)\;,\\
\fl	b(\eta)=\frac{b_{\rm max}}{12}\Bigg{[}\eta^2\left(1+\frac{\eta^2}{60}+\frac{\eta^4}{2520}\right)+ \frac{12\eta_\ast}{\eta}\left(1 - \frac{\eta^2}{12} - \frac{\eta^4}{720} - \frac{\eta^6}{30240}\right) + {\cal O}(\eta^7)\Bigg{]}~.
}
\end{eqnarray}

In the regime $\eta_\ast=0$ or $\eta_\ast/\eta\ll1$ such that the part dependent on $\eta_\ast$ from Eq.~\eref{solk1} drops out, we can work out the dependence of the scale factors on the cosmological time. Thus, for the relation between $\eta$ and cosmological time we have
\begin{eqnarray}
	\eqalign{
	\xi:=\frac{12\pi t}{t_{\rm max}}=\eta^3\left(1-\frac{\eta^2}{20}\right)+{\cal O}(\eta^7)\quad\textrm{and hence}\\
	\eta=\xi^{1/3}+\frac{\xi}{60}+{\cal O}(\xi^{5/3})
	}
\end{eqnarray}
so that
\begin{eqnarray}\label{solk1matterapp}
	\eqalign{
	a(\xi)=\frac{a_{\rm max}}{4}\xi^{2/3}\left(1-\frac{\xi^{2/3}}{20}\right)+{\cal O}(\xi^2)\;,\\
	b(\xi)=\frac{b_{\rm max}}{12}\xi^{2/3}\left(1+\frac{\xi^{2/3}}{20}\right)+{\cal O}(\xi^2)\;.
	}
\end{eqnarray}
Hence as long as
\begin{equation}
	\frac{\xi^{2/3}}{20}=\frac{1}{20}\left(\frac{12\pi t}{t_{\rm max}}\right)^{2/3}\simeq \left(\frac{t}{4t_{\rm max}}\right)^{2/3}\ll 1~,
\end{equation}
the $S^2$ and the $\chi$ direction are expanding with a scale factor proportional to $t^{2/3}$, the typical behaviour for stress energy from matter. Of course, as $t$ grows and approaches $t_{\rm max}$ the evolution of $a$ slows and eventually turns into contraction, while $b$ continues to grow. 

\vspace{4pt}
A quick calculation based on the approximate solution~\eref{solk1matterapp} shows that the two Hubble rates are
\begin{eqnarray}
	H_a=\frac{2}{3t}\left(1-\frac{\xi^{2/3}}{20}\right)\;,\quad
	H_b=\frac{2}{3t}\left(1+\frac{\xi^{2/3}}{20}\right)\quad
\end{eqnarray}
and hence
\begin{equation}
	t=\frac{4}{3(H_a+H_b)}\; .
\end{equation}
In particular, the age, $t_0$, of the universe is related to the present Hubble rates $H_{a,0}$ and $H_{b,0}$ by
\begin{equation}
	t_0=\frac{4}{3(H_{a,0}+H_{b,0})}~,\quad\mbox{as long as}\quad \frac{t_0}{4t_{\rm max}}\ll 1\; .
\end{equation}
Note, the age is determined by the mean $(H_{a,0}+H_{b,0})/2$ of the two Hubble rates.  
\vspace{4pt}

We find, to leading order in the $\eta$ expansion, for the ratio of shear to matter energy density
\begin{equation}
	\sqrt{\frac{\sigma^2/\kappa}{\rho}}\simeq \frac{2}{15}\frac{\bar{a}}{a_{\rm max}}\simeq \frac{\eta^2}{30}\; .
	\label{eq:shearbyexpansion}
\end{equation}
with the `average' scale factor $\bar{a}$ as defined in~\eref{Hbar}. As expected, for small anisotropy, $\eta\ll 1$, the shear is negligible.

\vspace{8pt}

\begin{figure}[htb]
	\centering
	\includegraphics[width=0.48\textwidth]{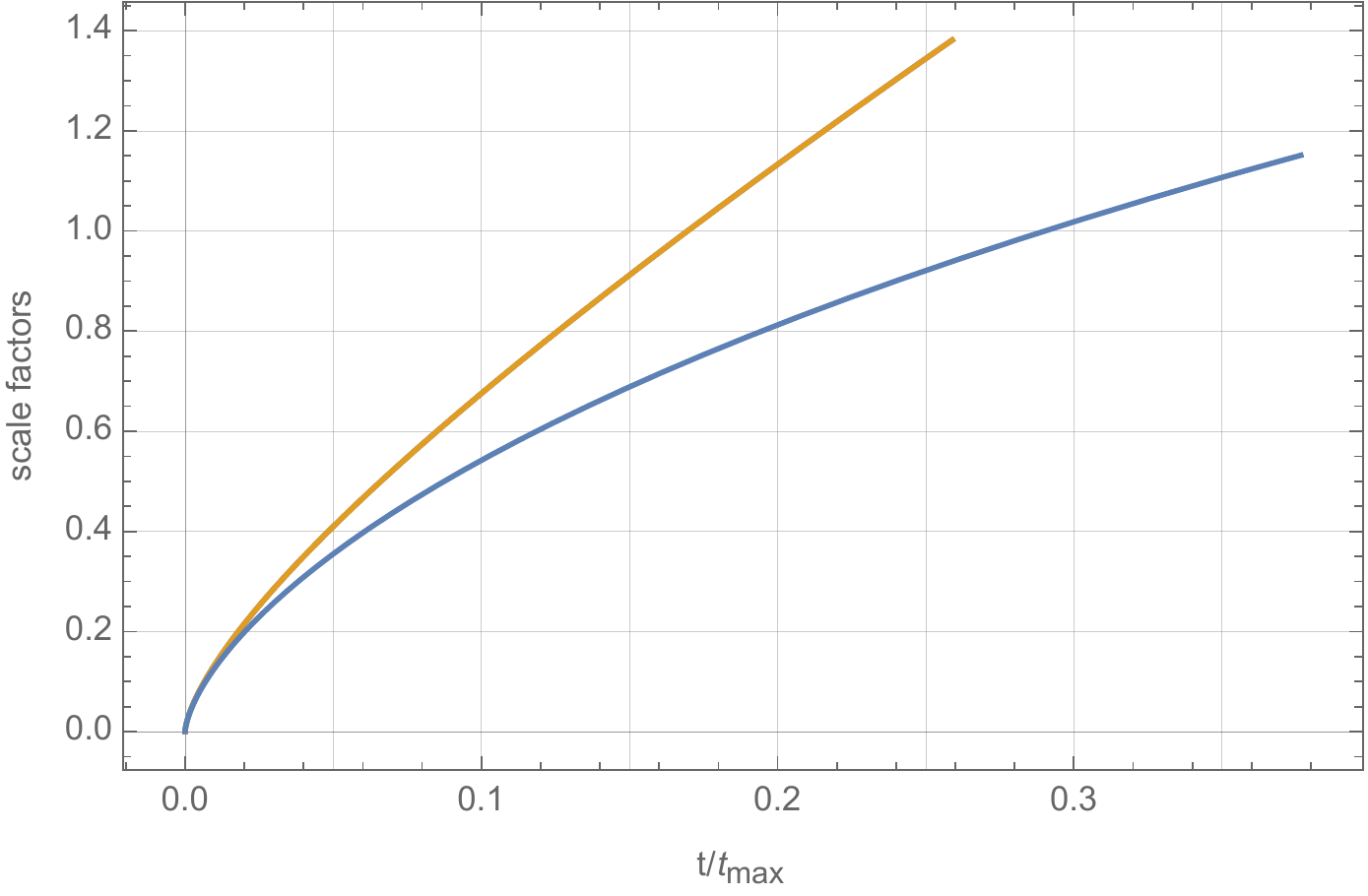}
	\caption{Scale factors in matter dominated open KS universe. The plot shows $a(t/t_{\rm max})/a_{\rm max}$ (orange curve, corresponding to the $H^2$ part of the metric) and $3b(t/t_{\rm max})/b_{\rm max}$ (blue curve, corresponding to the $X_1$ part of the metric). 
	By choosing $a_{\rm max}=b_{\rm max}/3$, the two scale factors have the same asymptotic form near $t=0$, $a(t)\simeq b(t)\sim t^{2/3}$.}
	\label{fig:MatterPlotOpen1}
\end{figure}

{\bfseries The open KS universe.} 
Formally, one can convert between the two sets of equations corresponding to the open/closed universes by $r_2\rightarrow i r_2$. This leads to the following negative curvature solutions
\begin{eqnarray}\label{solabk-1}
\eqalign{
	a(\eta)=e^{A(\eta)}=a_{\rm max}\sinh^2\left(\frac{\eta}{2}\right)\;,\\
	b(\eta)=e^{B(\eta)}=b_{\rm max}\left[\frac{\eta+\eta_\ast}{2}\coth\left(\frac{\eta}{2}\right)-1\right]\;.
	}
\end{eqnarray}
Integrating $dt=a(\tau)d\tau$ leads to the relation
\begin{equation}
	t=t_{\rm max}\left[\sinh(\eta)-\eta\right]\;,\qquad t_{\rm max}=\frac{a_{\rm max}{r_2}}{2}\; ,
\end{equation}
between co-moving time $t$ and $\tau$ (or $\eta=\tau/r_2$). Both $\eta$ and $t$ run in the interval $[0,\infty)$. The scale factor $a$ of $H_2$ vanishes at $\eta=0$ and increases forever.
 As before, we distinguish between two types of solutions that have an approximate FRLW limit, depending on the behaviour of $b$ near $\eta=0$: 
\vspace{4pt}

{\bfseries Type I solutions}: $\eta_\ast=0$. The universe is nearly isotropic at early times and the degree of anisotropy increases monotonically in time. With $a_{\rm max}=b_{\rm max}/3$, we have $a(t)\simeq b(t)\sim t^{2/3}$ at early times. The evolution of the scale factors is shown in \fref{fig:MatterPlotOpen1}.
\vspace{4pt}

\begin{figure}[htb]
	\centering
	\includegraphics[width=0.48\textwidth]{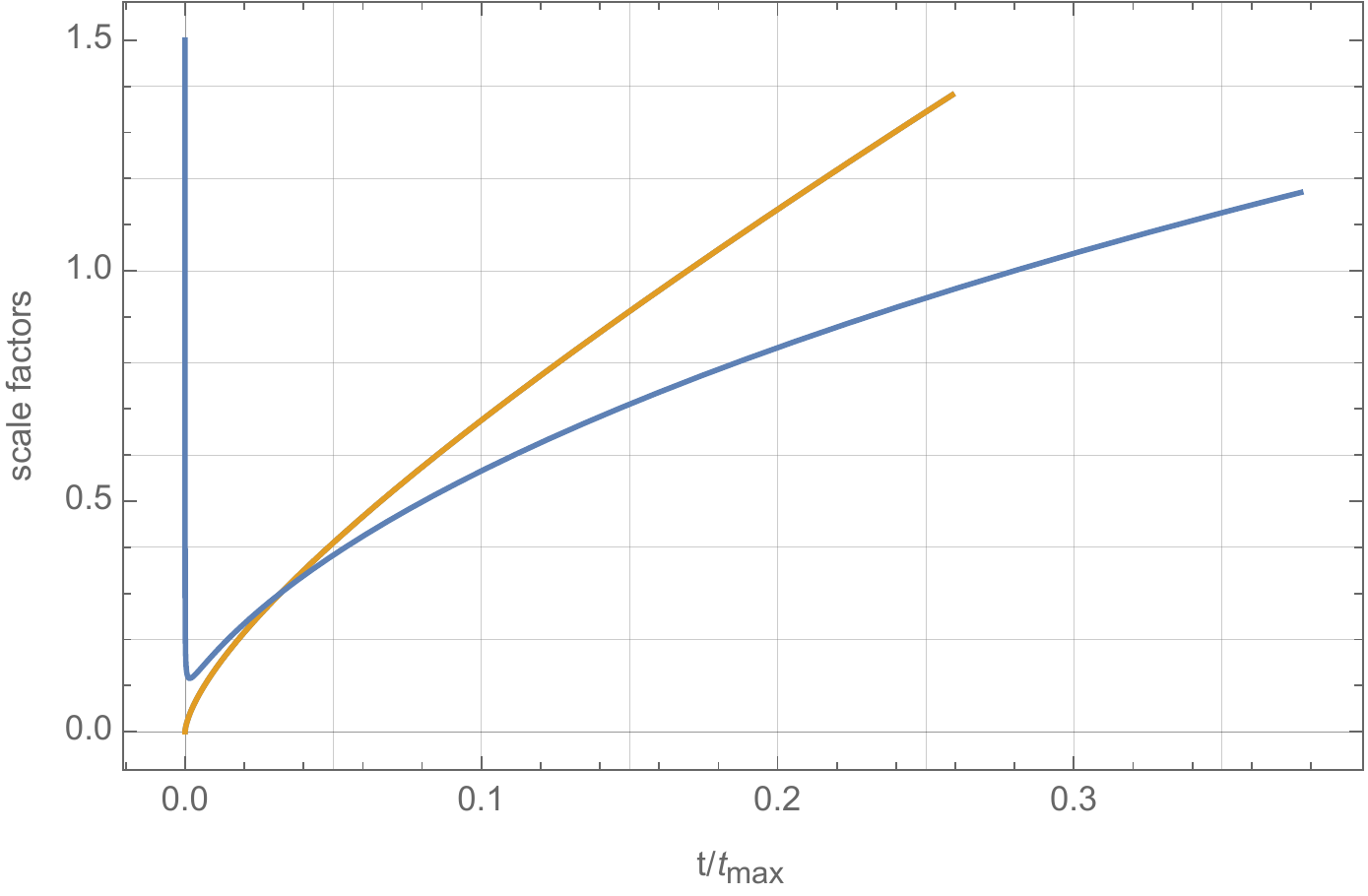} 	\hspace{10pt}
	\includegraphics[width=0.48\textwidth]{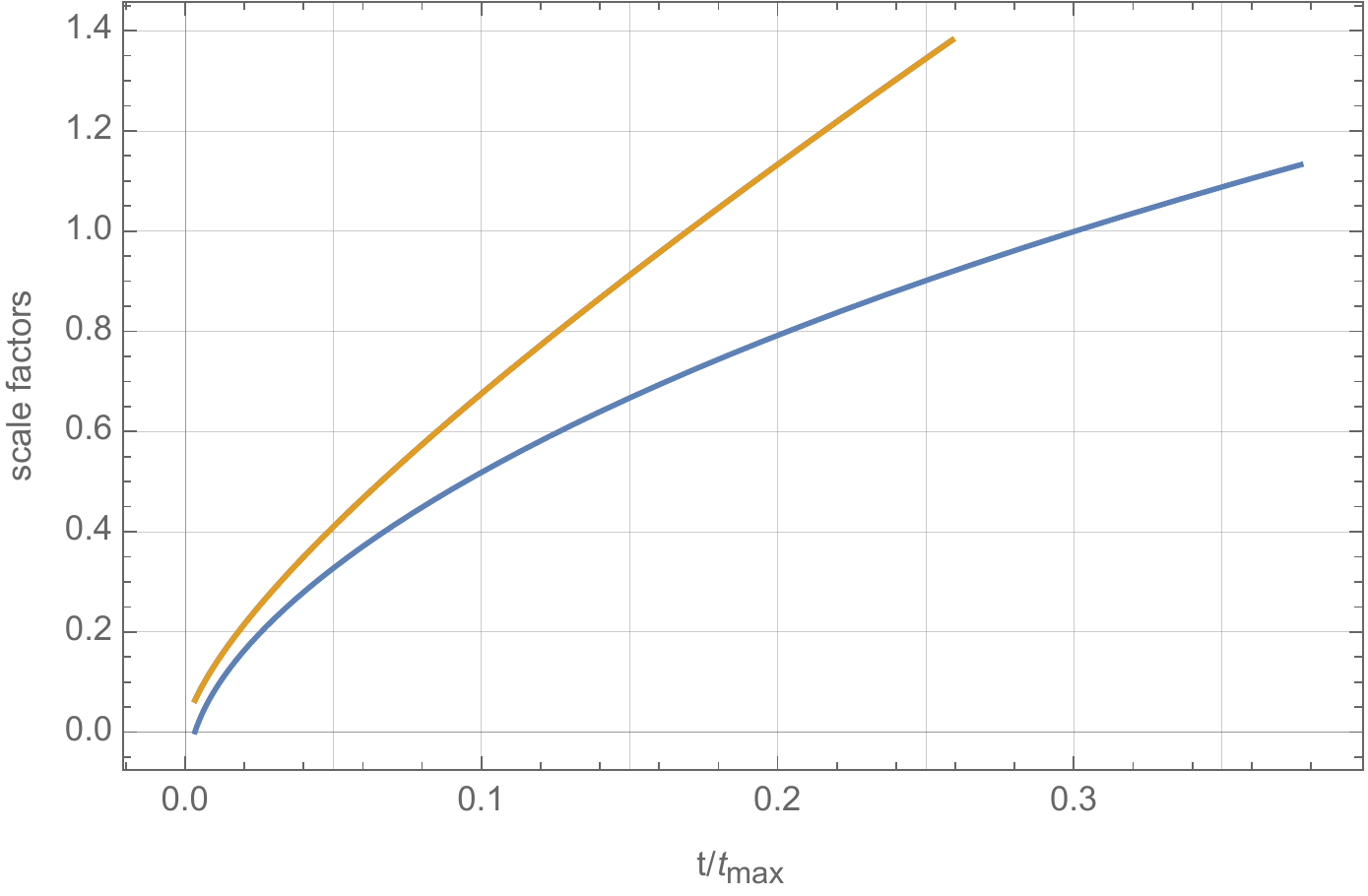}
	\caption{The evolution of the scale factors in matter dominated open KS universe when $\eta_*\neq0$. On the left $\eta_*=0.01$, while on the right $\eta_\ast=-0.01$. Both plots show $a(t/t_{\rm max})/a_{\rm max}$ (orange curve, corresponding to the $S^2$ part of the metric) and $3b(t/t_{\rm max})/b_{\rm max}$ (blue curve, corresponding to the $X_1$ part of the metric). The universe is nearly isotropic in the range $1\gg\eta\gg\eta_*$.}
	\label{fig:MatterPlotOpen2}
\end{figure}

{\bfseries Type II solutions}: $\eta_\ast\neq 0$. As in the closed KS case, when $\eta\gg |\eta_\ast|$, the evolution of $b$ is similar to the case $\eta_\ast=0$. However, for small values of $\eta$ the evolution depends on the sign of~$\eta_\ast$.   
 For $\eta_\ast>0$, the $X_1$ scale factor diverges at $\eta=0$. The cosmological evolution sees $b$ contracting from infinity to a minimal value and then expanding again. As in the closed KS case, for $\eta_\ast\ll1$, there is an intermediate regime, $\eta_\ast\ll\eta\ll 1$, in which the scale factors evolve approximately at the same rate. An illustration of this scenario is shown in the left hand side plot in \fref{fig:MatterPlotOpen2} for $\eta_*=0.01$. 
For $\eta_\ast<0$, $b$ goes to zero at a finite positive value of $\eta=\eta_{\rm min}$ and the range for $\eta$ should be accordingly restricted to $[\eta_{\rm min}, 2\pi]$. The universe starts in a two-dimensional phase. An illustration of this scenario is shown in the right hand side plot in  \fref{fig:MatterPlotOpen2} for $\eta_*=-0.01$.
\vspace{4pt}

The solutions \eref{solabk-1} can be expanded around $\eta=0$:
\begin{eqnarray}\label{solk-1}
\eqalign{
\fl	a(\eta)=\frac{a_{\rm max}}{4}\eta^2\left(1-\frac{\eta^2}{12}+\frac{\eta^4}{360}\right)+{\cal O}(\eta^8)\;,\\
\fl	b(\eta)=\frac{b_{\rm max}}{12}\Bigg{[}\eta^2\left(1-\frac{\eta^2}{60}+\frac{\eta^4}{2520}\right)+\frac{12\eta_\ast}{\eta}\left(-1 - \frac{\eta^2}{12} + \frac{\eta^4}{720} - \frac{\eta^6}{30240}\right) + {\cal O}(\eta^7)\Bigg{]}~.
	}
\end{eqnarray} 

We can work out the dependence of the scale factors on the cosmological time in the regime $\eta_\ast=0$ or $\eta_\ast/\eta\ll1$. The relation between $\eta$ and co-moving time is
\begin{eqnarray}
	\eqalign{
	\xi:=\frac{6 t}{t_{\rm max}}=\eta^3\left(1+\frac{\eta^2}{20}\right)+{\cal O}(\eta^7)\quad\textrm{and hence}\\
	\eta=\xi^{1/3}-\frac{\xi}{60}+{\cal O}(\xi^{5/3})~.
	}
\end{eqnarray}

Re-writing the above small-$\eta$ expressions for $a$ and $b$ in terms of co-moving time gives
\begin{eqnarray}
	\eqalign{
	a(\xi)=\frac{a_{\rm max}}{4}\xi^{2/3}\left(1+\frac{\xi^{2/3}}{20}\right)+{\cal O}(\xi^2)\;,\\
	b(\xi)=\frac{b_{\rm max}}{12}\xi^{2/3}\left(1-\frac{\xi^{2/3}}{20}\right)+{\cal O}(\xi^2)
	}
\end{eqnarray}
so an approximate expansion with the power $2/3$ for both scale factors as long as 
\begin{equation}
	\frac{\xi^{2/3}}{20}=\frac{1}{20}\left(\frac{6\,t}{t_{\rm max}}\right)^{2/3}=\left(\frac{3\,t}{20\sqrt{5}\,t_{\rm max}}\right)^{2/3}\ll 1~.
\end{equation}
Interpreted in the FLRW paradigm, such a universe would appear to be filled not only with matter, but with an additional fluid with $\omega_{\sigma} = -2/3$, similar to the case of the closed matter dominated KS universe above. 

\subsection{Solutions with matter and cosmological constant}
As before, we choose the gauge $N=A$ and introduce the dimensionless time coordinate $\eta=\tau/r_2$. The energy density and pressure are given by 
\begin{equation}
	\rho=\rho_M + \rho_\Lambda= \hat{\rho}e^{-2A-B}+\rho_\Lambda
\end{equation} 
and $p=-\rho_\Lambda$, with $\Lambda=\kappa\rho_\Lambda$. The Einstein equations, written in terms of $a(\eta)$ and $b(\eta)$, become
\begin{eqnarray}
	\eqalign{
	\frac{{a'}^2}{a^2}+2\frac{a'b'}{ab}=\frac{b_{\rm max}}{b}+\lambda \frac{a^2}{a_{\rm max}^2}-k\\
	2\frac{a''}{a}-\frac{{a'}^2}{a^2}=\lambda \frac{a^2}{a_{\rm max}^2}-k\\
	\frac{a''}{a}-\frac{{a'}^2}{a^2}+\frac{b''}{b}=\lambda \frac{a^2}{a_{\rm max}^2}
	}
\end{eqnarray} 
where the prime now denotes $d/d\eta$, $a_{\rm max}$ is an arbitrary value for the scale factor and 
\begin{equation}\label{bmaxlambda}
	b_{\rm max}:=\kappa\,\hat{\rho}\,r_2^2\;,\qquad \lambda=r_2^2a_{\rm max}^2\Lambda\; .
\end{equation}
These equations can be solved by starting with a power series ansatz for $a(\eta)$ and $b(\eta)$ of the form
\begin{eqnarray}
	\eqalign{
	a(\eta)=\alpha_0\eta^2+\alpha_1\eta^4+\alpha_2\eta^6+\alpha_3\eta^8+{\cal O}(\eta^{10})\;,\\
	b(\eta)=\beta_0\eta^2+\beta_1\eta^4+\beta_2\eta^6+\beta_3\eta^8+{\cal O}(\eta^{10})\; .
	}
\end{eqnarray}
The ansatz is motivated by the solutions obtained in the previous case of matter domination. Inserting this into the field equations and matching terms order by order in $\eta$ leads to the solutions
\begin{eqnarray}\label{eq:abcc}
	\eqalign{
	a(\eta)=\frac{a_{\rm max}\eta^2}{4}\left[1{-}\frac{k\eta^2}{12}{+}\frac{\eta^4}{360}{+}\left(\lambda{-}\frac{k}{15}\right)\frac{\eta^6}{1344}\right]{+}{\cal O}(\eta^{10})\\
	b(\eta)=\frac{b_{\rm max}\eta^2}{12}\left[1{+}\frac{k\eta^2}{60}{+}\frac{\eta^4}{2520}{+}\left(\lambda{+}\frac{k}{75}\right)\frac{\eta^6}{1344}\right]{+}{\cal O}(\eta^{10})
	}
\end{eqnarray}
The first three terms are exactly as in the solutions~\eref{solk1} without a cosmological constant which only comes in at the fourth order. The ratio $f=a/b$ will be required frequently and is given by
\begin{equation}\label{aoverbexp}
	f=\frac{a}{b}=\frac{3a_{\rm max}}{b_{\rm max}}\left[1-\frac{k\eta^2}{10}+\frac{17\eta^4}{4200}-\frac{11k\eta^6}{126000}+{\cal O}(\eta^8)\right]
\end{equation}
Interestingly, the cosmological constant dependence in the term proportional to $\eta^6$ drops out. The two associated scale factors are
\begin{eqnarray}
	\eqalign{
	H_a{=}\frac{8}{a_{\rm max}r_2\eta^3}\left[1{-}\frac{\eta^4}{240}{+}\frac{(9\lambda-2k)\eta^6}{6048}{+}{\cal O}(\eta^8)\right]\\
	H_b{=}\frac{8}{a_{\rm max}r_2\eta^3}\left[1{+}\frac{k\eta^2}{10}{+}\frac{17\eta^4}{2800}{+}\left(\lambda{+}\frac{221\,k}{1125}\right)\frac{\eta^6}{672}+{\cal O}(\eta^8)\right]
	}
\end{eqnarray}
Further, the co-moving time $t{=}\int_0^\tau d\tau\, a(\tau){=}r_2\int_0^\eta d\eta\, a(\eta)$ is given by
\begin{equation}
	t=\frac{r_2 a_{\rm max}\eta^3}{12}\left[1{-}\frac{k\eta^2}{20}{+}\frac{\eta^4}{840}{+}\left(\lambda{-}\frac{k}{15}\right)\frac{\eta^6}{4032}{+}{\cal O}(\eta^8)\right]\; .
\end{equation}

For the energy densities $\rho_{\rm M}$, $\rho_\Lambda$ and $\rho_k$ of matter, cosmological constant and curvature, respectively, this implies
\begin{eqnarray}
\fl	\rho_{\rm M}=\frac{\hat{\rho}}{a^2b}=\bar{\rho}\frac{a_{\rm max}^2b_{\rm max}}{a^2b}=\bar{\rho}\frac{192}{\eta^6}\left[1+\frac{3k\eta^2}{20}+\frac{13\eta^4}{1050}-\left(\lambda-\frac{1133\, k}{3375}\right)\frac{\eta^6}{448}+{\cal O}(\eta^8)\right]\nonumber\\
\fl	\rho_\Lambda=\frac{\Lambda}{8\pi G_N}=\bar{\rho}\lambda\nonumber\\
\fl	\rho_k=-\frac{k}{8\pi G_N r_2^2 a^2}=-k\bar{\rho}\frac{a_{\rm max}^2}{a^2}=-\bar{\rho}\frac{16\, k}{\eta^4}\left[1+\frac{k\eta^2}{6}+\frac{11\eta^4}{720}-\left(\lambda-\frac{31\,k}{45}\right)\frac{\lambda^6}{672}+{\cal O}(\eta^8)\right]\nonumber\\
\fl	\bar{\rho}:=\frac{1}{8\pi G_N r_2^2 a_{\rm max}^2}
\end{eqnarray}
and for the ratios of energy densities
\begin{eqnarray}
	\eqalign{
	\frac{\rho_\Lambda}{\rho_{\rm M}}=\frac{\lambda\eta^6}{192}\left[1{-}\frac{3k\eta^2}{20}{+}\frac{17\eta^4}{1680}{+}\left(\lambda{-}\frac{124\,k}{675}\right)\frac{\eta^6}{448}{+}{\cal O}(\eta^8)\right]\\
	\frac{\rho_k}{\rho_M}=-\frac{k\eta^2}{12}\left[1{+}\frac{k\eta^2}{60}{+}\frac{\eta^4}{2520}{+}\left(\lambda{+}\frac{k}{75}\right)\frac{\eta^6}{1344}{+}{\cal O}(\eta^8)\right]
	}
\end{eqnarray}

\begin{figure}[htb]
	\centering
	\includegraphics[width=0.48\textwidth]{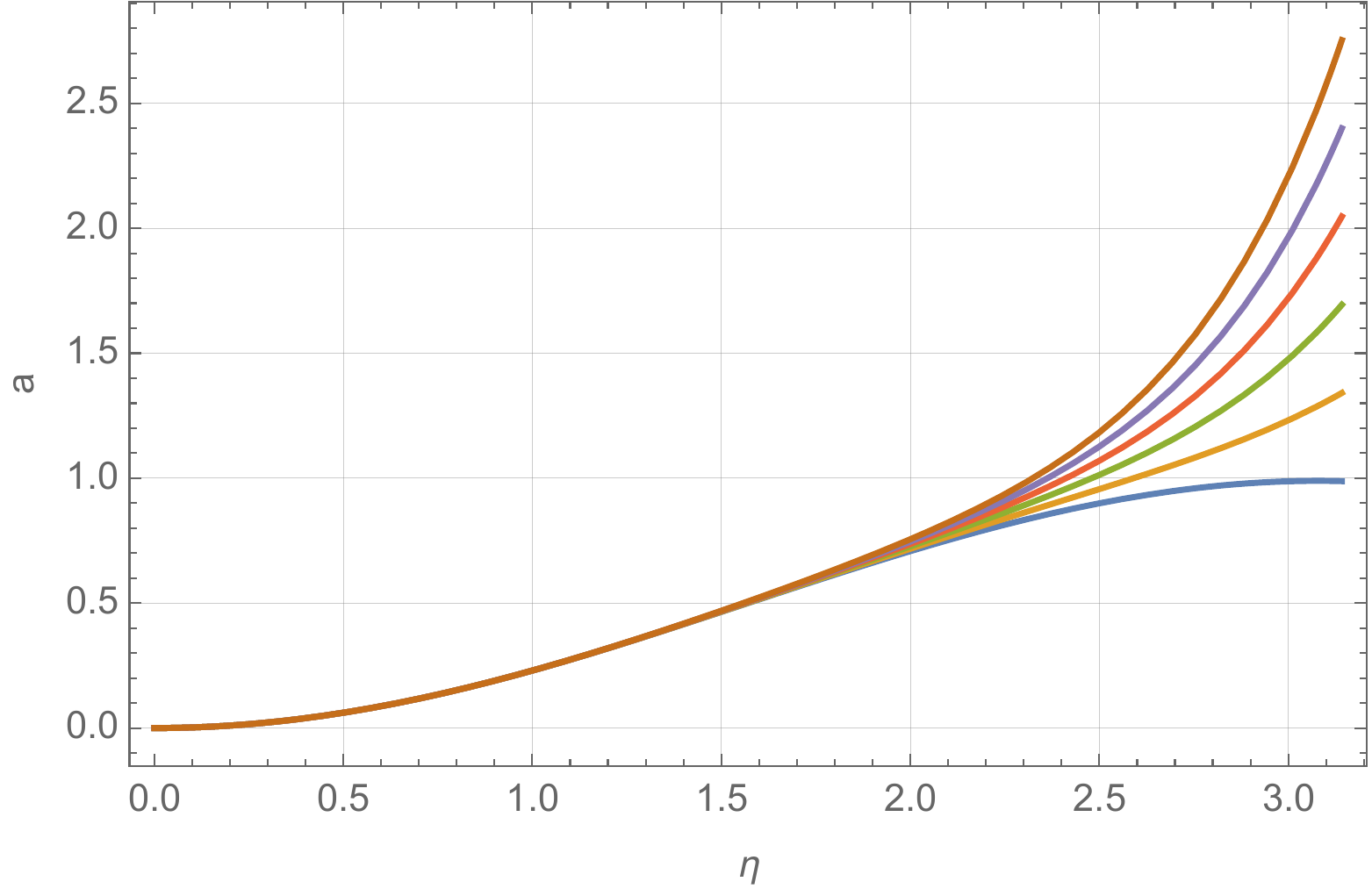}
	\includegraphics[width=0.48\textwidth]{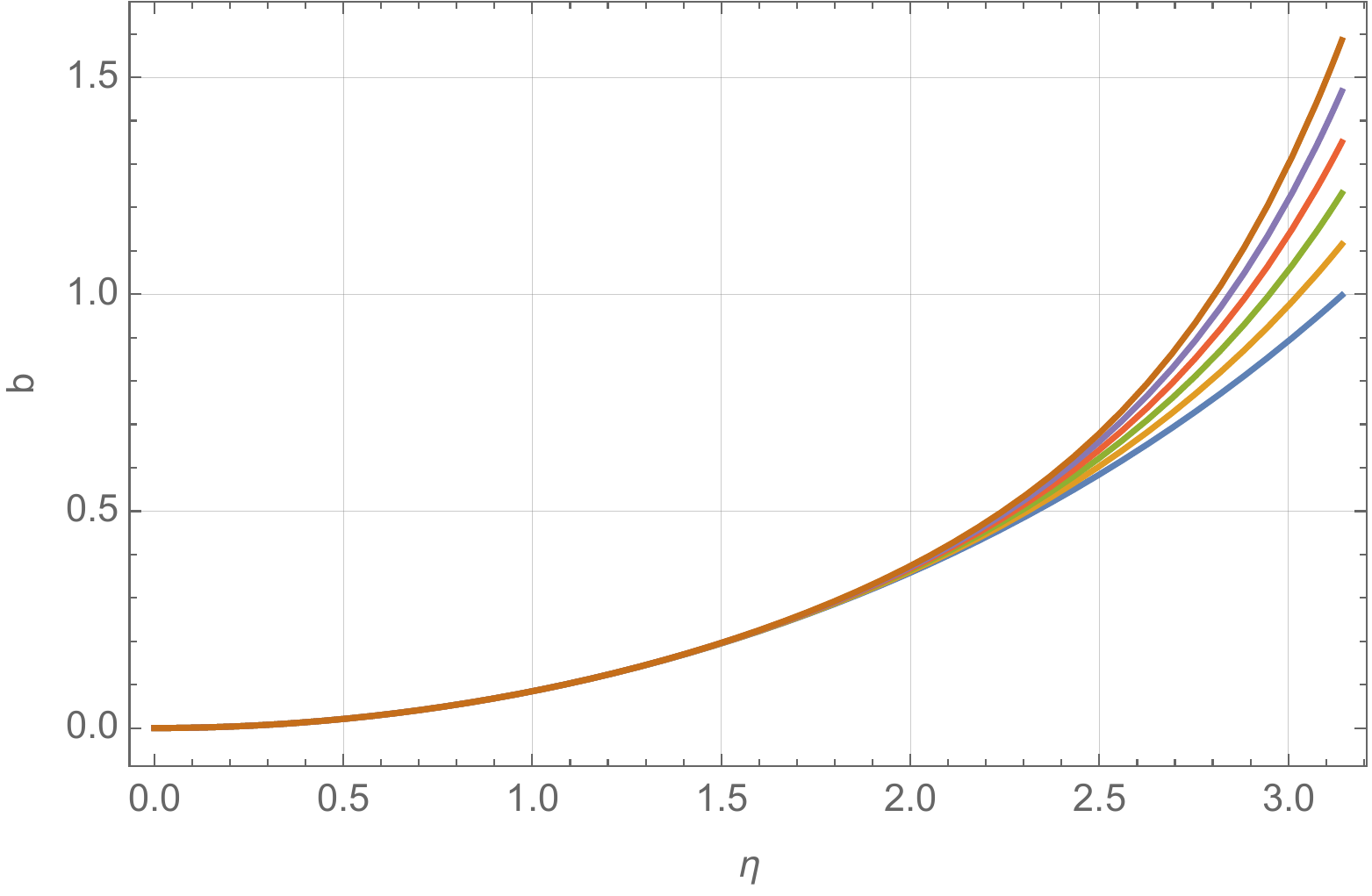}
	\caption{Left: evolution of $a(\eta)$ for different values of $\lambda$ and $a_{\rm max}=1$, calculated to $\mathcal O (\eta^{12})$ in the series expansion. Right: evolution of $b(\eta)$ for different values of $\lambda$ and $b_{\rm max}{=}1$, calculated to $\mathcal O (\eta^{12})$ in the series expansion. Values of $\lambda$ (from top to bottom): 1.0, 0.8, 0.6, 0.4, 0.2, 0.0. }
	\label{fig:MatterLambdaAB}
\end{figure}

For orientation, the scale factors $a$ and $b$ for the positive curvature case, $k=1$, are shown in \fref{fig:MatterLambdaAB} for values $\lambda=0,0.2,0.4,0.6,0.8,1$, and setting $a_{\rm max}=b_{\rm max}=1$. Note that in the presence of a cosmological constant the $\eta$ range is finite and the scale factors diverge as we approach the upper limit for $\eta$. The above approximation to order $\eta^6$ breaks down close to this point.

\subsection{Solutions with radiation}
In this case the equation of state is $p=\rho/3$ and the useful gauge is $N=-B$, so that the Einstein field equations~\eref{eqn:EFE} become

\begin{eqnarray}\label{einstein1rad}
	\eqalign{
	{A'}^2{+}2A'B'=\kappa\,\rho\, e^{-2B}{-}\frac{k}{r_2^2}e^{-2B-2A}\\
	2A''{+}3{A'}^2{+}2A'B'=-\frac{\kappa}{3}\,\rho\,e^{-2B}{-}\frac{k}{r_2^2}e^{-2B-2A}\\
	A''{+}B''{+}{A'}^2{+}2{B'}^2{+}2A'B'=-\frac{\kappa}{3}\, \rho\, e^{2N}
	}
\end{eqnarray}
with $\rho=\hat{\rho}\, e^{-8A/3-4B/3}$. 
\vspace{4pt}

{\bfseries The closed KS universe.} We discuss first the case $k=1$ case. Subtracting the second equation from the first and adding four multiples of the last equation gives 
\begin{equation}
	A''+2B''+{A'}^2+4{B'}^2+4A'B'=0\; .
\end{equation}
Introducing the linear combination $C=A+2B$, this becomes $C''+{C'}^2=0$ with solution
\begin{equation}
	C(\eta)=\ln(\eta)\;,\qquad \eta:=\frac{\tau-\tau_0}{\tau_3}\; ,
\end{equation}
where $\tau_0$ and $\tau_3$ are integration constants.  Replacing $2B=C-A$ in the first equation leads to an equation purely for $A$ with solution
\begin{equation}
	A(\eta)=\ln\left(\alpha_2\eta^{1/3}{-}\alpha_1\eta{+}\alpha_0\right)\;,~ \alpha_2{:=}3\kappa\,\rho_0\,\tau_3^2\;,~ \alpha_1{:=}\frac{\tau_3^2}{r_2^2}\; ,
\end{equation} 
and $\alpha_0$ is arbitrary. It follows that 
\begin{equation}\label{solk1rad}
	B(\eta){=}\frac{1}{2}(C(\eta){-}A(\eta)){=}\frac{1}{2}\ln(\eta){-}\frac{1}{2}\ln\left(\alpha_2\eta^{1/3}{-}\alpha_1\eta{+}\alpha_0\right)
\end{equation}
The co-moving time can be found from the relation $dt=\pm \tau_3 e^{-B}d\eta$ but integrating this directly leads to a very complicated result in terms of elliptical functions. This being case, it is better to keep the solutions in terms of $\tau$ and use numerical results for the relation to co-moving time, when needed. 

\begin{figure}[htb]
	\centering
	\includegraphics[width=0.48\textwidth]{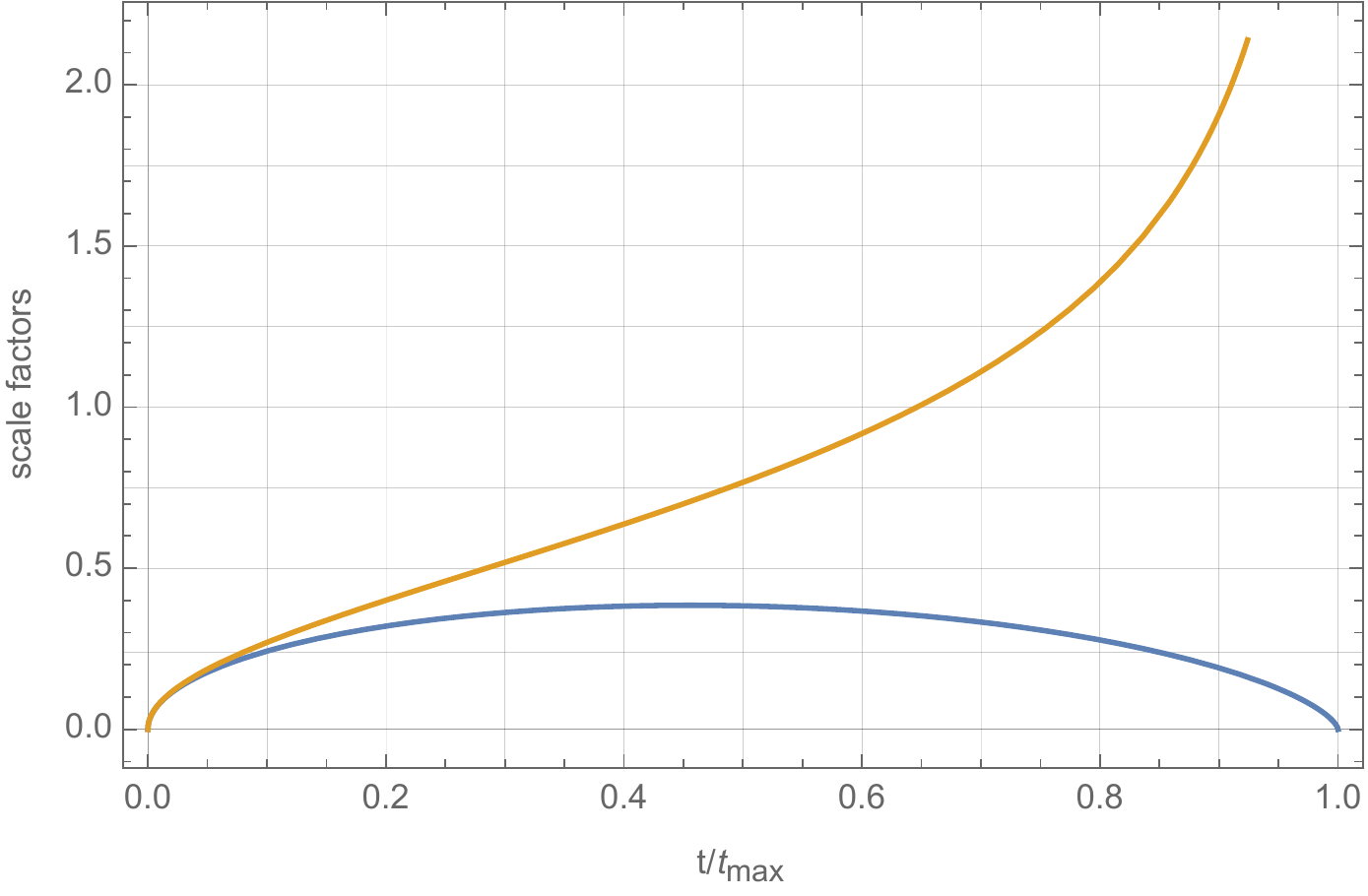}
	\caption{Evolution of the scale factors in the radiation dominated closed KS universe. The plot shows $a(t/t_{\rm max})/a_{i}$ (orange curve, corresponding to the $S^2$ part of the metric) and $b(t/t_{\rm max})/b_{i}$ (blue curve, corresponding to the $S^1$ part of the metric).}
	\label{fig:RadiationPlotClosed}
\end{figure}

Things simplify considerably if we assume the initial condition $a(\eta=0)=0$ which means that $\alpha_0=0$. In this case, the time parameter $\eta$ is in the range
\begin{equation}
	\eta\in [0,\eta_{\rm max}]\;,\qquad \eta_{\rm max}=\left(\frac{\alpha_2}{\alpha_1}\right)^{3/2}\; .
\end{equation} 
In this case, both scale factors $a=e^A$ and $b=e^B$ vanish at $\eta=0$, $a$ expands initially, then re-collapses and becomes zero at $\eta=\eta_{\rm max}$, while $b$ expands throughout and diverges at $\eta=\eta_{\rm max}$. With $\alpha_0=0$ we can also integrate $dt=\pm \tau_3 e^{-B}d\eta$ and this leads to
\begin{equation}
	1{-}\left(\frac{\eta}{\eta_{\rm max}}\right)^{2/3}{=}\,\left(1{-}\frac{t}{t_{\rm max}}\right)^{2/3}\!\!\!,~~
	t_{\rm max}=\frac{\tau_3\alpha_2^{3/2}}{\alpha_1}\; ,
\end{equation}
so that the interval $[0,\eta_{\rm max}]$ is mapped to the co-moving time interval $[0,t_{\rm max}]$. In terms of $\eta_{\rm max}$, the solution~\eref{solk1rad} can be re-written as
\begin{eqnarray}\label{ab_rad_closedKS}
	\eqalign{
	a(\eta)=e^{A(\eta)}=\frac{\alpha_2^{3/2}}{\alpha_1^{1/2}}\left(\frac{\eta}{\eta_{\rm max}}\right)^{1/3}\left(1-\left(\frac{\eta}{\eta_{\rm max}}\right)^{2/3}\right)\;,\\
	b(\eta)=e^{B(\eta)}=\alpha_1^{-1/2}\left(\frac{\eta}{\eta_{\rm max}}\right)^{1/3}\left(1-\left(\frac{\eta}{\eta_{\rm max}}\right)^{2/3}\right)^{-1/2}
	}
\end{eqnarray}
while in terms of the co-moving time, this becomes
\begin{eqnarray}
	\eqalign{
	a(t)=a_i\left(1-\left(1-\frac{t}{t_{\rm max}}\right)^{2/3}\right)^{1/2}\left(1-\frac{t}{t_{\rm max}}\right)^{2/3}\;,\\
	b(t)=b_i\left(1-\left(1-\frac{t}{t_{\rm max}}\right)^{2/3}\right)^{1/2}\left(1-\frac{t}{t_{\rm max}}\right)^{-1/3}\;,\\
	a_i=\frac{\alpha_2^{3/2}}{\alpha_1^{1/2}} \;,~~ t_{\rm max}=r_2 a_i\;,~~ b_i=(3\kappa\,\rho_0\,r_2^2)^{3/4}a_i^{-1/2}~.
	}
\end{eqnarray}
The evolution of the scale factors is shown in \fref{fig:RadiationPlotClosed}.

Setting $\epsilon=t/t_{\rm max}$ we can study the $\epsilon\ll 1$ early-time limit of this solution which is
\begin{eqnarray}
	\eqalign{
	a(\epsilon)=a_i\sqrt{\frac{2}{3}}\epsilon^{1/2}\left( 1 -\frac{7\epsilon}{12}+{\cal O}(\epsilon^{2})\right)~,\\
	b(\epsilon)=b_i\sqrt{\frac{2}{3}}\epsilon^{1/2}\left( 1 +\frac{5\epsilon}{12}+{\cal O}(\epsilon^{2})\right)~.
	}
\end{eqnarray}
To leading order, this is a power law with exponent $1/2$ for both scale factors, as expected for radiation in an isotropic universe.
Keeping the first two terms in the expansion, the Hubble rates are
\begin{equation}
	H_a=\frac{1}{2t}- \frac{7}{12t_{\rm max}} \;,\quad
	H_b=\frac{1}{2t}+ \frac{5}{12t_{\rm max}}~.
\end{equation}
Interpreted in the FLRW paradigm, this universe appears to be filled not only with radiation, but also with a fluid with equation of state $p=-\rho$ and energy density
\begin{equation}
\frac{\sigma^2}{3} = \frac{(H_a-H_b)^2}{9} \simeq \frac{1}{9t_{\rm max}^2} \ll \frac{\kappa \rho_{\rm rad}}{3} = \frac{1}{4t^2}~.
\end{equation}

\vspace{8pt}
{\bfseries The open KS universe.} The Einstein equations differ from the ones for positive curvature only by the sign of the curvature term and they can be obtained from the $k=1$ equations by the formal replacement $r_2\rightarrow i r_2$. This means the solutions is now given by~\eref{solk1rad} with $r_2\rightarrow i r_2$, that is
\begin{eqnarray}\label{solk-1rad}
	\eqalign{
	A(\eta){=}\ln\left(\alpha_2\eta^{1/3}{+}\alpha_1\eta{+}\alpha_0\right)\\
	B(\eta){=}\frac{1}{2}(C(\eta){-}A(\eta)){=}\frac{1}{2}\ln(\eta){-}\frac{1}{2}\ln\left(\alpha_2\eta^{1/3}{+}\alpha_1\eta{+}\alpha_0\right)\\
	\eta:=\frac{\tau-\tau_0}{\tau_3}\;,\quad \alpha_2:=3\,\kappa\,\hat{\rho}\,\tau_3^2\;,\quad \alpha_1:=\frac{\tau_3^2}{r_2^2}\; .
	}
\end{eqnarray}
As before, working out co-moving time by inegrating the relation $dt=\pm\tau_3\, e^{-B} d\eta$ leads to a complicated result involving elliptic functions. Adopting again the initial condition $a(\eta=0)=0$, that is, $\alpha_0=0$, this simplifies significantly and leads to
\begin{equation}
	\left(1{+}\frac{t}{t_1}\right)^{2/3}{=}\,1{+}\left(\frac{\eta}{\eta_1}\right)^{2/3}\!\!\!, ~ \eta_1{=}\left(\frac{\alpha_2}{\alpha_1}\right)^{3/2}\!\!\!,~
	t_1{=}\frac{\tau_3\alpha_2^{3/2}}{\alpha_1}\; ,
\end{equation}
and both $\eta$ and co-moving time $t$ now run in the range $[0,\infty)$. In more concise terms, the solution~\eref{solk-1rad} for $\alpha_0=0$ can now be written as
\begin{eqnarray}\label{ab_rad_openKS}
	\eqalign{
	a(\eta)=e^{A(\eta)}=a_i\left(\frac{\eta}{\eta_{1}}\right)^{1/3}\left(1+\left(\frac{\eta}{\eta_{1}}\right)^{2/3}\right)\;,\\
	b(\eta)=e^{B(\eta)}=b_i\left(\frac{\eta}{\eta_{1}}\right)^{1/3}\left(1+\left(\frac{\eta}{\eta_{1}}\right)^{2/3}\right)^{-1/2}\;,\\
	a_i:=\frac{\alpha_2^{3/2}}{\alpha_1^{1/2}}\;,\qquad b_i:=\alpha_1^{-1/2}~.
	}
\end{eqnarray}
In terms of co-moving time this becomes
\begin{eqnarray}
	\eqalign{
	a(t)=a_i\left(\left(1+\frac{t}{t_1}\right)^{2/3}-1\right)^{1/2}\left(1+\frac{t}{t_1}\right)^{2/3}\\
	b(t)=b_i\left(\left(1+\frac{t}{t_1}\right)^{2/3}-1\right)^{1/2}\left(1+\frac{t}{t_1}\right)^{-1/3}~\\
	t_{1}=r_2 a_i\;,\qquad b_i=(3\,\kappa\,\rho_0\,r_2^2)^{3/4}a_i^{-1/2}~.
	}
\end{eqnarray}

\begin{figure}[htb]
	\centering
	\includegraphics[width=0.48\textwidth]{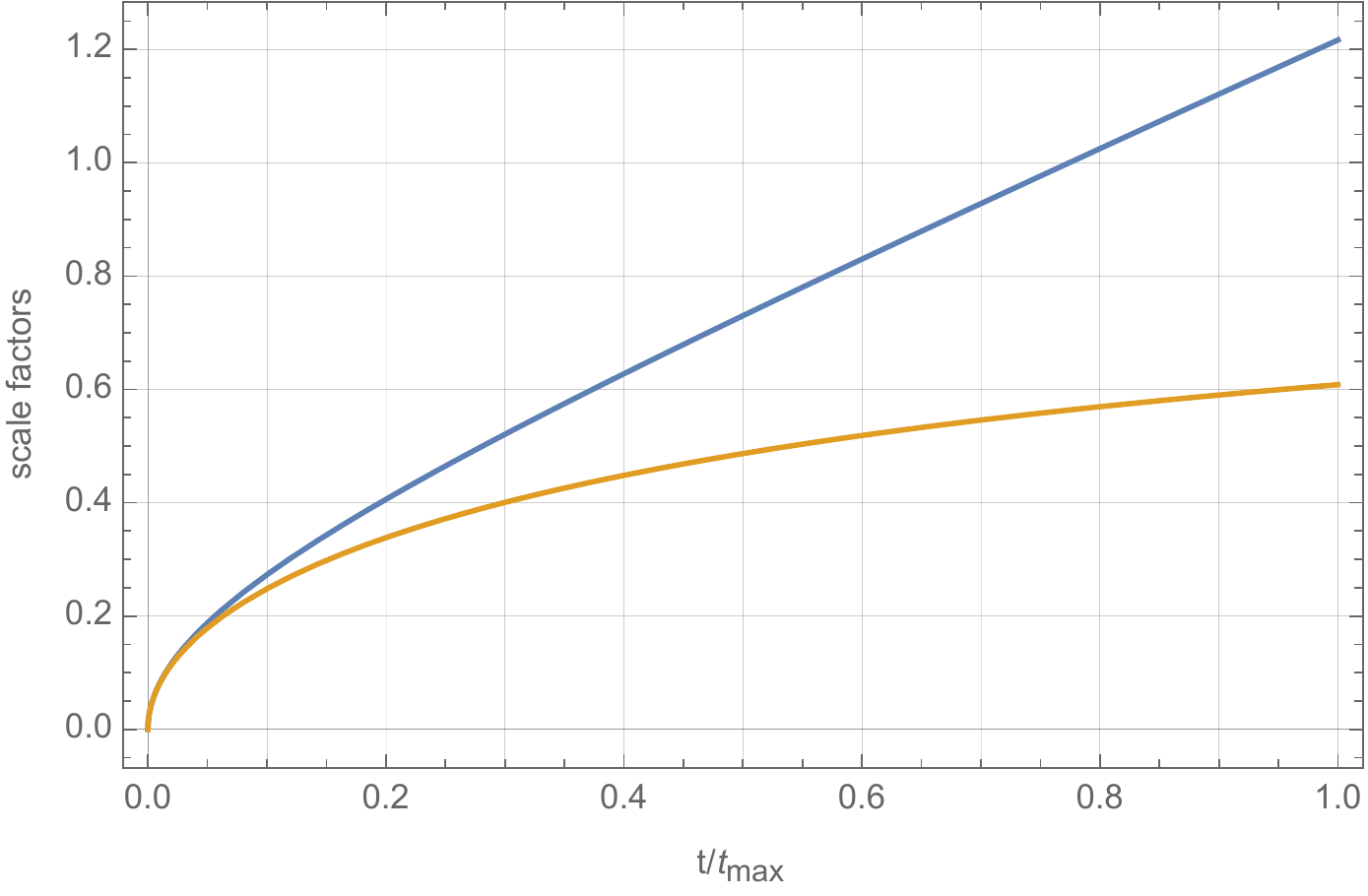}
	\caption{Evolution of the scale factors in the radiation dominated open KS universe. The plot shows $a(t/t_{\rm max})/a_{i}$ (orange curve, corresponding to the $H^2$ part of the metric) and $b(t/t_{\rm max})/b_{i}$ (blue curve, corresponding to the $S^1$ part of the metric).}
	\label{fig:RadiationPlotOpen}
\end{figure}

The evolution of the scale factors is shown in \fref{fig:RadiationPlotOpen}.

For $\epsilon=t/t_{1}$ the early-time limit, $\epsilon\ll 1$, becomes
\begin{eqnarray}
	\eqalign{
	a(\epsilon)=a_i\left[\sqrt{\frac{2}{3}}\epsilon^{1/2}+\frac{7}{6\sqrt{6}}\epsilon^{3/2}+{\cal O}(\epsilon^{5/2})\right]\;,\\
	b(\epsilon)=b_i\left[\sqrt{\frac{2}{3}}\epsilon^{1/2}-\frac{5}{6\sqrt{6}}\epsilon^{3/2}+{\cal O}(\epsilon^{5/2})\right]~.
	}
\end{eqnarray}
As before, to leading order these are power laws with exponent $1/2$, as it is characteristic for radiation. 

\subsection{Solutions with more general perfect fluids}
The solutions $a(\eta)$ and $b(\eta)$ obtained in the case of a radiation-filled universe (Eqns.~\eref{ab_rad_closedKS} for the closed KS universe and Eqns.~\eref{ab_rad_openKS} for the open KS universe) can be directly rendered as solutions for the case of a perfect fluid with equation of state $p=w\rho$ with $w>0$ or $w<-1$, provided that we choose the lapse function as
\begin{equation}
N = -B +\frac{1}{2}\log\frac{1+1/w}{4}~.
\end{equation}

\section{Propagation of Light}\label{sec:light_prop}

Drawing any quantitative conclusions from the anisotropic cosmologies discussed above requires an understanding of how light propagates in such backgrounds. Similar studies on the propagation of light in anisotropic and homogeneous universes have been present in the literature for several decades (see, for instance the early~\cite{Tomita_68, Saunders_69} or the more recent~\cite{Schucker:2014wca, Fleury:2014rea, Awwad:2022uoz}), demonstrating that anisotropy can lead to a cosmological lensing effect. 
In the following sections we will derive explicit formulae for the red-shift, the angular diameter distance and the luminosity distance as measured by an observer in a closed or open KS universe.

\subsection{Redshift}\label{sec:Redshift}
Consider a co-moving observer with four-velocity given by $U_\mu=(n,0,0,0)$, such that $g^{\mu\nu}U_\mu U_\nu=-1$, and a photon propagating along a null-geodesic 
\begin{equation}
X^\mu(s)=(\tau(s),\chi(s),\theta(s),\phi(s))
\end{equation} 
with tangent vector $V^\mu(s)=dX^\mu/ds$. As discussed in~\ref{app:Geodesic}, setting the initial condition $V^\phi(s_{\rm in})=0$, the geodesic equation implies that the photon propagates along the corresponding contant-$\phi$ curve. The remaining components of $V^\mu(s)$ are given by
\begin{eqnarray}
	V^{\mu}(s){=} \left(\pm\sqrt{\frac{r_1^2 v_1^2}{b(s)^2 n(s)^2}{+}\frac{r_2^2v_2^2}{a(s)^2 n(s)^2}},  ~\frac{v_1}{b(s)^2}, ~\frac{v_2}{a(s)^2},~ 0\right)
\end{eqnarray}
where $v_1$ and $v_2$ are constants of integration. 
 
The photon frequency $\omega$, as measured by the co-moving observer, is then
\begin{equation}\label{redshift_1}
	\omega=V^\mu U_\mu=\sqrt{\frac{r_1^2 v_1^2}{b^2}+\frac{r_2^2v_2^2}{a^2}}~.
\end{equation}
In particular, for photons propagating in the $\chi$ and $\phi$ directions we have, as expected,
\begin{equation}
	\omega=	\left\{\begin{array}
				{cl}\displaystyle\frac{r_1 v_1}{b} &\mbox{for photons in }\chi\mbox{ direction }(v_2=0)~,\\[12pt]
				\displaystyle\frac{r_2 v_2}{a} &\mbox{for photons in }\theta\mbox{ direction }(v_1=0)~.
               		\end{array}\right. 
\end{equation}

The above formulae can be recast in a simpler form by introducing the angle measured in the $(\chi,\theta)$-plane between the $\theta$-direction, corresponding to the vector $e_\theta=(0,1)^T$, and the tangent vector $V=(V^\chi,V^\theta)$. This  angle can be computed with the spatial part of the metric
\begin{equation}\label{hatg0}
	\hat{g}:=r_1^2b^2 d\chi^2+r_2^2a^2 d\theta^2\; ,
\end{equation}
which then gives
\begin{eqnarray}
	\eqalign{
	\cos(\sphericalangle(V,e_\theta))&=\frac{\hat{g}(V,e_\theta)}{\sqrt{\hat{g}(V,V)}\sqrt{\hat{g}(e_\theta,e_\theta)}}=\frac{r_2 v_2 b}{\sqrt{r_1^2 v_1^2 a^2 +r_2^2 v_2^2 b^2 }}=:\cos(\alpha)
	}
\end{eqnarray}

In particular, the angle $\alpha_0$ between the incoming ray and the $\theta$-direction as measured at the present time $\tau_0$ is
\begin{equation}\label{alphadef1}
	\tan(\alpha_0)=\frac{r_1v_1a_0}{r_2v_2b_0}\; ,
\end{equation}
where $a_0=a(\tau_0)$ and $b_0=b(\tau_0)$. With this notation, the redshift $z$ of a photon propagating from an early time $\tau$ with scale factors $a=a(\tau)$, $b=b(\tau)$ to the present time $\tau_0$ is given by
\begin{eqnarray}\label{redshift}
	\eqalign{
	1{+}z(\tau,\tau_0,\alpha_0)=\frac{\omega(\tau)}{\omega_0}&=\sqrt{\sin^2(\alpha_0)\frac{b_0^2}{b(\tau)^2}{+}\cos^2(\alpha_0)\frac{a_0^2}{a(\tau)^2}}\\
	&=\frac{a_0}{a(\tau)}(1{+}F_z(\tau,\tau_0,\alpha_0))\;,\\
	1 + F_z(\tau,\tau_0,\alpha_0)&:=\sqrt{\cos^2(\alpha_0){+}\sin^2(\alpha_0) (f(\tau)/f_0)^{2}}
	}
\end{eqnarray}   
where $f_0=a_0/b_0$ and $f(\tau)=a(\tau)/b(\tau)$.\\

\begin{figure}[htb]
	\centering
	\includegraphics[width=0.48\textwidth]{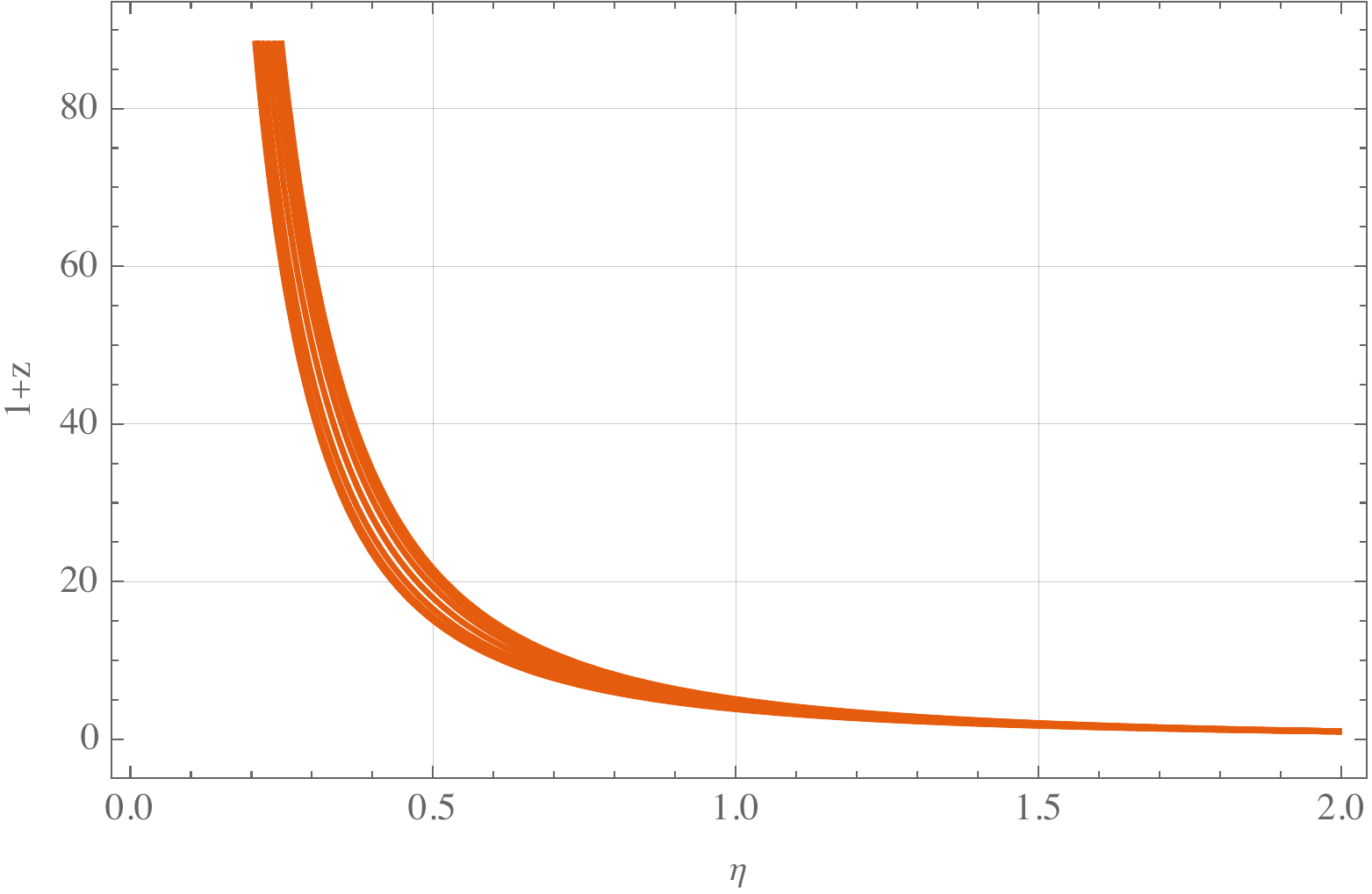}
	\caption{Plot of the redshift $1+z(\eta,\eta_0, \alpha_0)$ in matter-dominated closed KS universe, for $\eta_0=2$ and $10$ equally spaced values of $\alpha_0$ between $0$ and $\pi/2$. The lower curve corresponds to $\alpha_0=0$ and the upper curve corresponds to  $\alpha_0=\pi/2$.}
	\label{fig:RedShift}
\end{figure}

We illustrate the effect of the anisotropic background on the redshift in the case of the closed matter-dominated KS universe. Recall that in this case it is useful to set $n(\tau)=a(\tau)$ and to parametrize time in terms of the variable $\eta=\tau/r_2$. We first consider solutions of type I, in which the universe is isotropic at early times and becomes increasingly anisotropic at later times. To lowest order in $\eta$, we have $a(\eta)=b(\eta)$, $a_0=b_0$, so that $f=f_0=1$ and $F_z=0$. The redshift formula becomes
\begin{equation}
	1+z^{(0)}(\eta,\eta_0,\alpha_0)=\frac{a_0}{a(\eta)}\; ,
\end{equation}
independent of the angle $\alpha_0$. This, of course, is the standard FLRW result.
To the next order in $\eta$, we have
\begin{equation}\label{fres}
	f(\eta)=\frac{a(\eta)}{b(\eta)}=1-\frac{k\eta^2}{10}+{\cal O}(\eta^4)\; .
\end{equation}
where we have set $b_{\rm max}=3a_{\rm max}$ to simplify expressions (since we still keep the reference sizes $r_1$ and $r_2$ arbitrary, this is not actually a restriction of the solution).  Then it follows that 
\begin{equation}\label{Fzres}
	F_z(\eta,\eta_0,\alpha_0)=\frac{k\eta_0^2}{10}(1-\zeta^2)\sin^2(\alpha_0)\;,\qquad \zeta:=\frac{\eta}{\eta_0} 
\end{equation}
so that 
\begin{equation}
	1{+}z^{(1)}(\eta,\eta_0,\alpha_0)=\frac{a_0}{a}\left[1+\frac{k\eta_0^2}{10}(1{-}\zeta^2)\sin^2(\alpha_0)\right]~.
\end{equation}
The deviation from the FLRW result varies with the angle $\alpha_0$ and is proportional to $\eta_0^2$. 
Since the CMB temperature anisotropies are of $10^{-5}$ amplitude (after subtracting the dipole), this limits the range for $\eta_0$ to roughly $\eta_0\leq 10^{-2}$. Consequently, the variation of the redshift with the angle $\alpha_0$ is very small.
However, for values of $\eta_0$ close to unity, a significant variation of the redshift with the angle $\alpha_0$ can be achieved. Figure \ref{fig:RedShift} shows this variation for $\eta_0=2$. For photons propagating from a fixed, small value of $\eta$, the variation of the redshift $z(\eta,\eta_0,\alpha_0)$ with $\alpha_0$ is significant. 
\vspace{8pt}

\begin{figure}[htb]
	\centering
	\includegraphics[width=0.48\textwidth]{ScaleFactorsMatter3.pdf}\hspace{10pt}
	\includegraphics[width=0.48\textwidth]{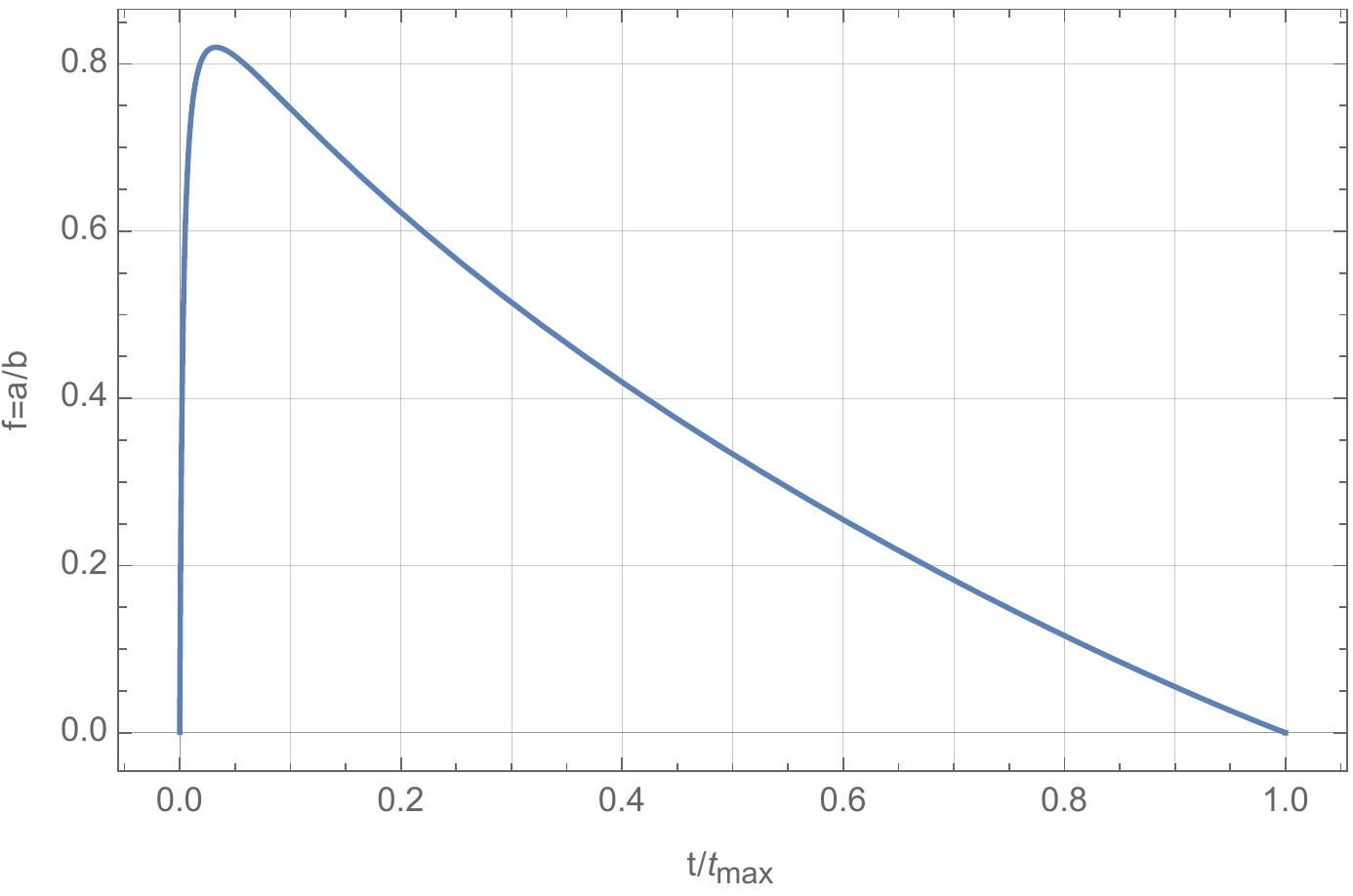}
	\caption{Evolution of the scale factors (left: orange curve shows $a(t)$, blue curve shows $b(t)$) and evolution of the ratio $f(t)=a(t)/b(t)$ (right) for the matter-dominated closed KS universe with $\eta_\ast=0.01$. The shape of $f(t)$ allows for a `coincidence' scenario where $f(t_{\rm rec})=f(t_0)$.}
	\label{f_closedKSplus}
\end{figure}

\begin{figure}[htb]
	\centering
	\includegraphics[width=0.48\textwidth]{ScaleFactorsMatterOpen4.pdf}\hspace{10pt}
	\includegraphics[width=0.48\textwidth]{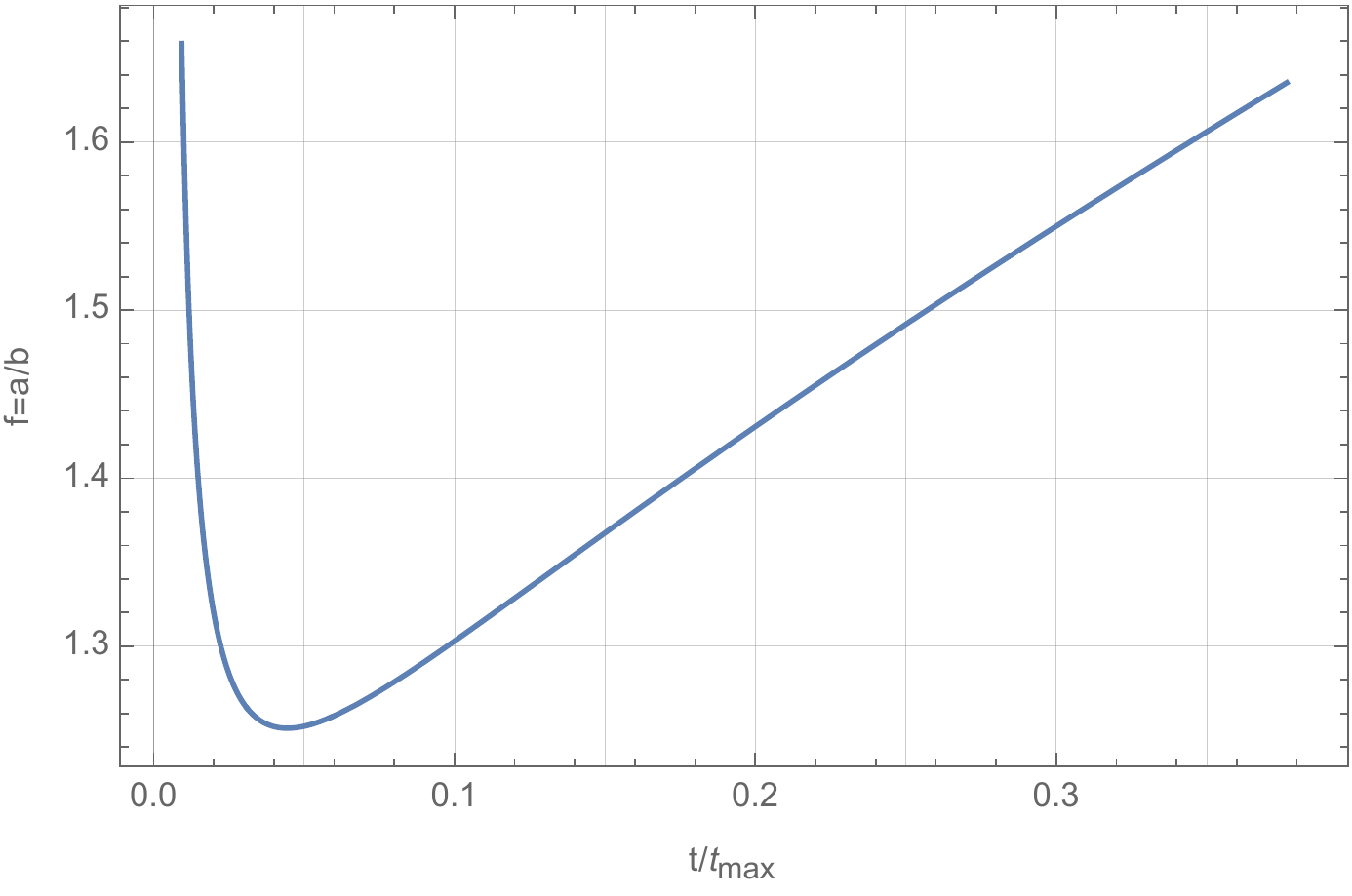}
	\caption{Evolution of the scale factors (left: orange curve shows $a(t)$, blue curve shows $b(t)$) and evolution of the ratio $f(t)=a(t)/b(t)$ (right) for the matter-dominated open KS universe with $\eta_\ast=-0.01$. The shape of $f(t)$ allows for a `coincidence' scenario where $f(t_{\rm rec})=f(t_0)$.}
	\label{f_openKSminus}
\end{figure}

An interesting possibility arises in the case of type II solutions, relying on the observation that the angular dependence in the redshift formula \eref{redshift} drops out provided that $f(t_{\rm rec})=f(t_0)$, where $t_{\rm rec}$ is the recombination time. Such a coincidence can be arranged when the integration constant $\eta_\ast$ is set to a non-zero value in the matter domination solutions for both the closed and the open KS universes. A plot of the ratio $f(t)=a(t)/b(t)$ is shown in figure~\ref{f_closedKSplus} for the closed KS universe and in figure~\ref{f_openKSminus} for the open KS universe. Of course, this scenario needs further investigation, which we defer to future work. 

\subsection{Angular Diameter Distance}\label{sec:add}
The angular diameter distance is defined by considering the angular sizes of objects in the night sky. In an FRLW background, given an object that fills out an angle $\delta\theta$ in the sky, while being extended over a proper distance~$D$, the angular diameter distance $d_A$ is given by 
\begin{equation}\label{dAdefinition}
	d_A = \frac{D}{\delta\theta}~.
\end{equation} 
In the FRLW setting (more generally, in any isotropic cosmology), the orientation of the arc along which the angular diameter distance is measured is irrelevant. This, of course, is no longer true in the anisotropic setting.

The effect of the anisotropic background on the angular diameter distance can be demonstrated by considering different orientations of the arc. In this section we look at two extreme cases: (1) the arc and the direction of sight are contained in a plane of constant $\phi$ and (2) the arc is perpendicular to the plane of constant $\phi$. In both cases the direction of sight makes an angle $\alpha_0$ with the $\theta$-direction and light travels along geodesics of constant~$\phi$. In the first case the angular diameter distance will be sensitive to the evolution of both scale factors, while in the second case it will only depend on the evolution of the $a$-scale factor. A general formula for the angular diameter distance for an arbitrary orientation of the arc can also be derived following a similar procedure as described below. 

In~\ref{app:AngDis} we present a detailed discussion of the angular diameter distance computation. Here we collect the relevant information.
For case (1), consider two photons with parameters $\alpha_0$ and $\tilde\alpha_0 = \alpha_0 + \delta\alpha_0$ which end up at the same location $(\chi_0,\theta_0)$ at time $\eta_0$, having been emitted at time $\tau_1$ from $(\chi_1,\theta_1)$ and $(\tilde\chi_1,\tilde\theta_1)=(\chi_1+\delta\chi_1,\theta_1+\delta\theta_1)$, respectively. The two geodesics are contained within the same $(\chi,\theta)$-plane. 

Null geodesics in the $(\chi,\theta)$-plane are described by the equations
\begin{eqnarray}\label{theta_chi_eq}
	\eqalign{
	\theta_f-\theta_i&=\pm \frac{1}{r_2}\int_{\tau_i}^{\tau_f} d\tau\,\frac{ n(\tau) /a(\tau)}{\sqrt{1+ \tan (\alpha(\tau))^2}}\\
	 \chi_f-\chi_i=& \pm \frac{1}{r_1}\int_{\tau_i}^{\tau_f} d\tau\,\frac{\tan (\alpha(\tau))\, n(\tau) /b(\tau)}{\sqrt{1+ \tan (\alpha(\tau))^2}}~,\\
	}
\end{eqnarray}
where
\begin{equation}\label{alphadef2}
	\tan(\alpha(\tau))=\frac{r_1v_1a(\tau)}{r_2v_2b(\tau)}=\frac{r_1v_1}{r_2v_2}f(\tau)\; 
\end{equation}
and $\alpha(\tau)$ is the angle between the tangent vector at time $\tau$ and the $\theta$-direction. It follows that 
\begin{equation}\label{tanalpha}
	\tan(\alpha(\tau))=\tan(\alpha_0)f(\tau)/f_0~,
\end{equation}
hence we can write
\begin{eqnarray}
	\eqalign{
	\theta(\tau, \alpha_0)  &{=}\theta_0{\pm}\frac{1}{r_2}\int_{\tau_0}^{\tau} d\tau'\,\frac{  n(\tau') /a(\tau')}{\sqrt{1+ \tan (\alpha_0)^2 (f(\tau')/f_0)^2 } }\\
	\chi(\tau, \alpha_0)  &{=}\chi_0{\pm}\frac{1}{r_1}\int_{\tau_0}^{\tau} d\tau'\,\frac{  \tan (\alpha_0) (f(\tau')/f_0){\cdot}n(\tau') /b(\tau')}{\sqrt{1+ \tan (\alpha_0)^2 (f(\tau')/f_0)^2 } }~.
	}
\end{eqnarray}

The physical extension of the source $D^{(\alpha)}$ is given by
\begin{eqnarray}\label{Ddef_main}
	\eqalign{
	D^{(\alpha)}(\tau_1,\tau_0,\alpha_0) &= \sqrt{r_1^2 b(\tau_1)^2 \delta\chi_1^2 + r_2^2 a(\tau_1)^2 \delta\theta_1^2}\\
	&=~\delta\alpha_0\sqrt{r_1^2 b(\tau_1)^2 \left.\frac{\partial\chi}{\partial\alpha_0}\right|_{\tau=\tau_1}\!\!\!\!\!\!\!\!  + r_2^2 a(\tau_1)^2 \left.\frac{\partial\theta}{\partial\alpha_0}\right|_{\tau=\tau_1}\!\!\!\!\!\!\!\!}~,
	}
\end{eqnarray}
where the partial derivatives are obtained by inserting the cosmological solutions of interest in the above expressions for $\theta$ and $\chi$. The superscript $(\alpha)$ indicates that the arc along which the distance is measured corresponds only to a variation of the angle $\alpha$. The angular diameter distance is then obtained as the ratio
\begin{equation}
	d_A^{(\alpha)}(\tau_1, \tau_0,\alpha_0) = \frac{D^{(\alpha)}(\tau_1,\tau_0,\alpha_0)}{\delta\alpha_0}~,
\end{equation}
which is a function of the time $\tau_1$ and, implicitly, for a given angle $\alpha_0$, a function of the redshift through~\eref{redshift}. 
\vspace{4pt}

We now turn to case (2) where the arc is oriented perpendicular to the plane of constant $\phi$ and, as before, the direction of sight makes an angle $\alpha_0$ with the $\theta$-direction and light travels along geodesics of constant~$\phi$. The arc corresponds to an angle $\delta\phi \cos(\alpha_0)$ measured by the observer. The physical extension of the source is 
\begin{equation}\label{Dphidef}
	D^{(\phi)}(\tau_1,\tau_0,\alpha_0) = a(\tau_1)r_2\, {\rm si}(\theta(\tau_1))\delta\phi~.
\end{equation}
Since the two rays travel along geodesics of constant $\phi$, the only way in which they can meet at the observer is that $\theta(\tau_0)=0$, which fixes the value of $\theta(\tau_1)$ through~\eref{theta_chi_eq}.
The corresponding angular diameter distance is then
\begin{equation}
	d_A^{(\phi)}(\tau_1, \tau_0,\alpha_0) = \frac{D^{(\phi)}(\tau_1,\tau_0,\alpha_0)}{\cos(\alpha_0)\delta\phi}~.
\end{equation}

The angle $\alpha_0$ is in principle measurable, assuming that the observer has knowledge of how the maximally symmetric two-dimensional subspace is embedded in three dimensions. The value of the present time $\tau_0$ is a measure of the current degree of anisotropy in the universe and has to be derived from astrophysical observations. The time $\tau_1$ is then obtained from the redshift via~\eref{redshift}.

\subsection{Luminosity Distance}\label{sec:LumDis}

The luminosity distance is defined through the flux-luminosity relationship,
\begin{equation}
	f  =\frac{L}{4\pi d_L^2}~.
\end{equation}
In flat-space, $d_L$ is simply the space distance between the source and the observer. In curved space this is not necessarily true. The flux measured by a detector of area $\Delta A$ coming from a bundle of rays emitted under a solid angle $\Delta\Omega$ is
\begin{equation}
	f  =\frac{L \Delta\Omega/(4\pi)}{\Delta A} \frac{1}{(1+z)^2}~,
\end{equation}
where the two factors of $(1+z)^{-1}$ account for the redshift in frequency and for the change in the rate of arrival of photons. The luminosity distance then takes the form
\begin{equation}
	d_L  =\sqrt{\frac{\Delta A}{\Delta \Omega}} (1+z)~.
\end{equation}

The name of the game is to compute the cross sectional area $\Delta A$ at time $\tau_0$ for a bundle of rays emitted at time $\tau_1$ under a solid angle $\Delta\Omega$. But first we need to choose an appropriate set of spherical coordinates centered around the source place at $(\chi_1,\theta_1)=(0,0)$. The local spatial metric in a neighbourhood of $(\chi_1,\theta_1)=(0,0)$ can be written as
\begin{eqnarray}
	\eqalign{
	\hat{g}_1&~=~b_1^2r_1^2d\chi^2+a_1^2r_2^2(d\theta^2+{\rm si}^2(\theta)d\phi^2)\\
	&\stackrel{\theta\ll1}{\simeq}b_1^2r_1^2d\chi^2+a_1^2r_2^2(d\theta^2+\theta^2d\phi^2)~,
	}
\end{eqnarray}
where ${\rm si}(x)=\sin(x)$ for $k=1$ and ${\rm si}(x)=\sinh(x)$ for $k=-1$, $a_1=a(\tau_1)$ and $b_1=b(\tau_1)$.
Introducing re-scaled coordinates ${\cal Z}=b_1 r_1\chi$ and ${\cal R}=a_1 r_2\theta$ this metric becomes the flat-space metric in cylindrical coordinates, which can be re-written in terms of spherical coordinates
\begin{eqnarray}
	\eqalign{
	\hat{g}_1& =d{\cal Z}^2+d{\cal R}^2+{\cal R}^2d\phi^2=dr^2+r^2(d\vartheta^2+\sin^2(\vartheta)d\phi^2)\\
	r&=\sqrt{{\cal R}^2+{\cal Z}^2},\; \cos(\vartheta)=\frac{\cal Z}{r},\; \sin(\vartheta)=\frac{\cal R}{r}~.
	}
\end{eqnarray}

In these spherical coordinates, we consider a bundle of null-geodesics between $\vartheta$, $\tilde{\vartheta}$ and $\phi$, $\tilde{\phi}$, with widths $\delta\vartheta=\tilde{\vartheta}-\vartheta$ and $\delta\phi=\tilde{\phi}-\phi$. This bunch intersects a sphere with radius $r$ around the origin in an area $r^2\sin(\vartheta)\delta\vartheta\,\delta\phi$. Given the total luminosity $L$ of the source, the energy flow $f_1$ through this area on the sphere is given by
\begin{equation}\label{f1}
	f_1=\frac{Lr^2\sin(\vartheta)\delta\vartheta\,\delta\phi}{4\pi r^2}=\frac{L\sin(\vartheta)\delta\vartheta\,\delta\phi}{4\pi}=\frac{L\cos(\alpha)\delta\alpha\,\delta\phi}{4\pi}\; ,
\end{equation}
where the last equality follows from $\vartheta=\frac{\pi}{2}-\alpha$, a relation which holds at $\vartheta=0$, as can be checked by transforming between the two frames. 

\vspace{4pt}
Since $f_1=L\Delta\Omega/(4\pi)$, it follows that 
\begin{equation}\label{dLformula}
	d_L = \sqrt{\frac{D^{(\alpha)} D^{(\phi)}}{\cos(\alpha) \delta\alpha\delta\phi}} (1+z)~,
\end{equation}
where $D^{(\alpha)}$ and $D^{(\phi)}$ correspond to the physical extension of the bundle in the $\alpha$- and, respectively, $\phi$-direction, as measured at $\tau_0$ (present time). For the size $D^{(\alpha)}$ we write a similar formula as in the computation \eref{Ddef_main} for the angular diameter distance performed in the previous section 
\begin{eqnarray}
	\eqalign{
	D^{(\alpha)} &= \sqrt{r_1^2 b(\tau_0)^2 \delta\chi_0^2 + r_2^2 a(\tau_0)^2 \delta\theta_0^2}\\
	&=~\delta\alpha\sqrt{r_1^2 b(\tau_0)^2 \left.\frac{\partial\chi}{\partial\alpha}\right|_{\tau=\tau_0}\!\!\!\!\!\!\!\!  + r_2^2 a(\tau_0)^2 \left.\frac{\partial\theta}{\partial\alpha}\right|_{\tau=\tau_0}\!\!\!\!\!\!\!\!}~.
	}
\end{eqnarray}
In the notation of~\eref{Ddef_main}, this is $D^{(\alpha)}(\tau_0,\tau_1,\alpha_1)$, i.e.~the role of the source and observer are swapped. The angle $\alpha_1=\alpha(\tau_1)$ is not measurable, however it can be related to $\alpha_0$ via~\eref{tanalpha}.

The size $D^{(\phi)}$ of the bunch in the $\phi$-direction is given by
\begin{equation}
	D^{(\phi)} = a_0r_2{\rm si}(\theta_0)\delta\phi\; ,
\end{equation}
while the redshift can be computed using~\eref{redshift}. 
\vspace{4pt}

To compare with empirical data for the redshift vs.~luminosity distance plots, we need to evaluate $H_0 d_L$, where today's Hubble rate $H_0$ can be taken as the averaged value $H_0=(2H_{a,0}+H_{b,0})/3$. Recall from Section~\ref{sec:EE} that different notions of averaging for the Hubble parameter can be considered, however, the qualitative conclusions of the discussion below remain the same. For a matter-dominated KS universe, today's Hubble rate $H_{a,0}$ for the scale factor $a$ is given by
\begin{eqnarray}
	\eqalign{
	H_{a,0}&=\left.\frac{1}{a_0}\frac{da}{d\eta}\frac{1}{dt/d\eta}\right|_{\eta=\eta_0}=\frac{1}{r_2a_{\rm max}}\frac{{\rm co}(\eta_0/2)}{{\rm si}^3(\eta_0/2)},
	}
\end{eqnarray}
where ${\rm si}(x)$ was defined above, while ${\rm co}(x)=\cos(x)$ for $k=1$ and ${\rm co}(x)=\cosh(x)$ for $k=-1$. On the other hand, today's Hubble rate $H_{b,0}$ for the scale factor $b$ is given by
\begin{eqnarray}
	\eqalign{
	H_{b,0}&=\left.\frac{1}{b_0}\frac{db}{d\eta}\frac{1}{dt/d\eta}\right|_{\eta=\eta_0}=\frac{1}{r_2a_{\rm max}}\frac{{\rm si(\eta_0)-\eta_0}}{{\rm si}^2(\eta_0/2)(\eta_0{\rm si}(\eta_0)+2{\rm co}(\eta_0)-2)}~.
	}
\end{eqnarray}

To first non-trivial order in $\eta$ expansion, the luminosity distance is given by
\begin{equation}\label{dLangle}
	d_L^{(1)}=d_L^{(0)}\left(1+\frac{k\eta_0^2}{40}(h_0(z)+h_2(z)\sin^2(\alpha_0))\right)
\end{equation}
where the zeroth order result $d_L^{(0)}$ is given by the flat, matter-dominated FRLW universe result, 
\begin{equation}\label{H0dLFRW}
	H_0 d_L^{(0)}= 2(1+z)\left(1-\frac{1}{\sqrt{1+z}}\right)\; ,
\end{equation}
and the functions $h_0(z)$ and $h_2(z)$ are of order $1$ for $1\lesssim z\lesssim 10$.

Of course, this formula provides a good approximation only for $\eta\ll 1$. Note from~\eref{eq:abcc} that the cosmological constant enters the expression for $d_L$ at order $\lambda\eta^6$ and, while this term can be large, it is $\alpha$-independent and, hence, does not contribute to the anisotropy in $d_L$. 

For reference, we also give the luminosity distance formula valid in the $\Lambda$CDM context:
\begin{equation}
	H_0 d_L^{\Lambda{\rm CDM}} (z) = (1+z)\int_{0}^z \frac{dt}{\sqrt{\Omega_m(1+t)^3+\Omega_\Lambda}}~,
\end{equation}
where $\Omega_\Lambda=0.75$ and $\Omega_m=0.25$.

Finally, we remark that the usual Etherington reciprocity relation between the angular diameter distance and the luminosity distance does not hold, in general, in anisotropic universes, given the definitions~\eref{dAdefinition} and~\eref{dLformula}. The reason is simple: as discussed in Section~\ref{sec:add}, the expression for the angular diameter distance depends on the orientation of the arc along which it is measured. Our formula in~\eref{dLformula} can be considered as a {\itshape generalized Etherington reciprocity relation} and would schematically read
\begin{equation}
	d_L = \sqrt{d_A^{(\alpha)}d_A^{(\phi)}} (1+z)~.
\end{equation}
More precisely
\begin{eqnarray}
	\eqalign{
	d_L(\tau_1,\tau_0,\alpha_0) = &\sqrt{d_A^{(\alpha)}(\tau_0,\tau_1,\alpha_1)\,d_A^{(\phi)}(\tau_0,\tau_1,\alpha_1)} \\ 
	&~~~~~~~~~~~~\times(1+z(\tau_1,\tau_0,\alpha_0))~
	}
\end{eqnarray}
where $\alpha_1$ and $\alpha_0$ are related by~\eref{tanalpha}.
For a generalized Etherington reciprocity relation valid for more general (anisotropic) backgrounds, under the same definitions of $d_A$ and $d_L$, see~\cite{Awwad:2022uoz}.

\section{Outlook on observational constraints}\label{sec:observables}

In this section we provide some qualitative remarks on the expected observational signatures of the open and closed KS universes as well as relations to the existing constraints. 

\subsection{Luminosity distance versus redshift}
\begin{figure}[htb]
	\centering
	\includegraphics[width=0.447\textwidth]{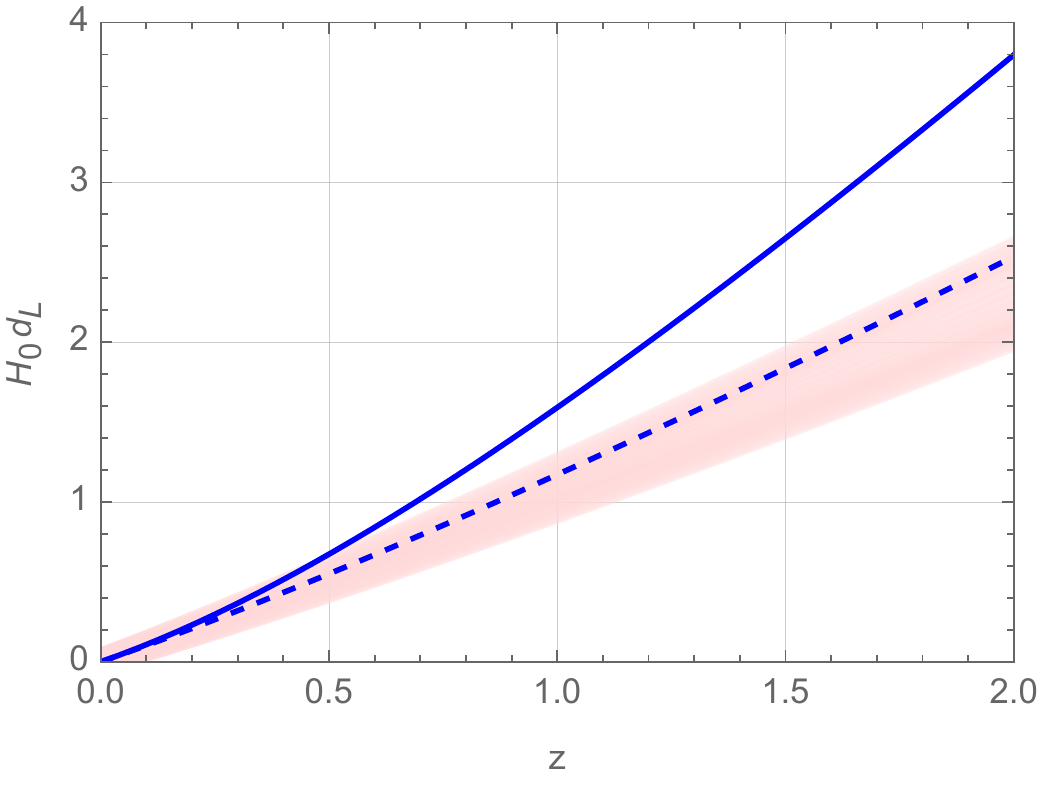}
	\includegraphics[width=0.58\textwidth]{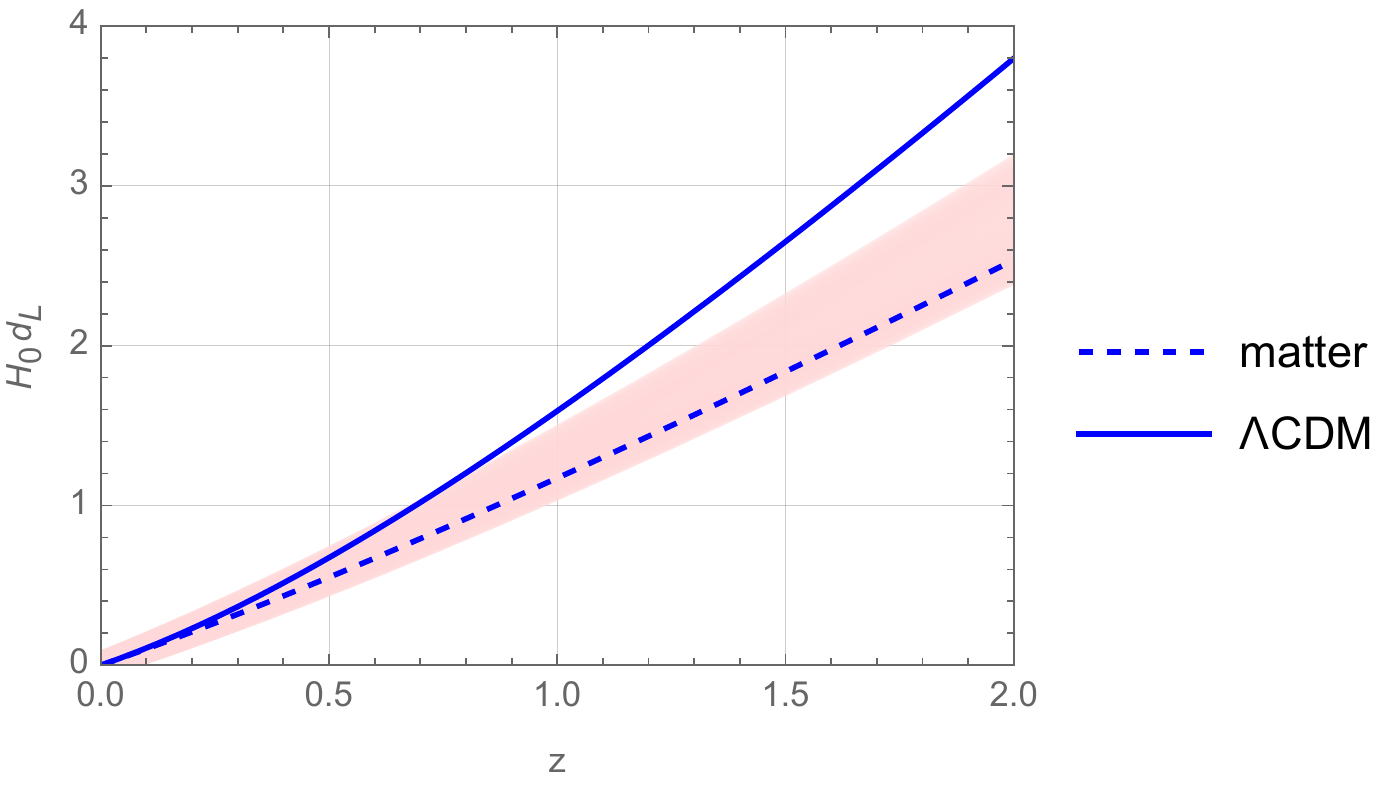}
	\caption{Parametric plots of luminosity distance $H_0d_L$ as a function of $z$ for the closed KS model (left) and for the open KS model (right) with $\eta_0=1.5$.  The pink bands indicate the variation with $\alpha_0$. For comparison, the plots also contain the result for a flat FLRW universe with matter (blue dashed curve) and for a flat FLRW universe with 25\% matter and 75\% cosmological constant (blue curve). 
	} 
	\label{fig:LD1}
\end{figure}

Our earlier discussion around~\eref{rho_ghost} suggests that, due to the presence of shear, the anisotropy of the background might lead to effects similar to that of a cosmological constant. To investigate this further, we consider our models with a vanishing cosmological constant, $\lambda=0$, and plot the luminosity distance $d_L$ as a function of redshift~$z$. 
In this context the value $\eta_0$ of the conformal time today should be interpreted as a measure of the anisotropy of the model. The luminosity distance curves for a moderately large value $\eta_0=1.5$, for the open and closed case are shown in \fref{fig:LD1}. 

For small $\eta_0\ll 1$ our models lead to a $d_L$ curve approximately equal to the FLRW matter curve (the dashed blue curves in the above plots), as~\eref{dLangle} shows. For the larger value, $\eta_0=1.5$, the $d_L$ curves for the closed KS universe move `in the wrong direction', further away from the $\Lambda$CDM curve (the blue curve in \fref{fig:LD1}). However, for the open KS model, the modification is in the right direction, bringing the $d_L$ curve for our model closer to the $\Lambda$CDM one.

\begin{figure}[htb]
	\centering
	\includegraphics[width=0.58\textwidth]{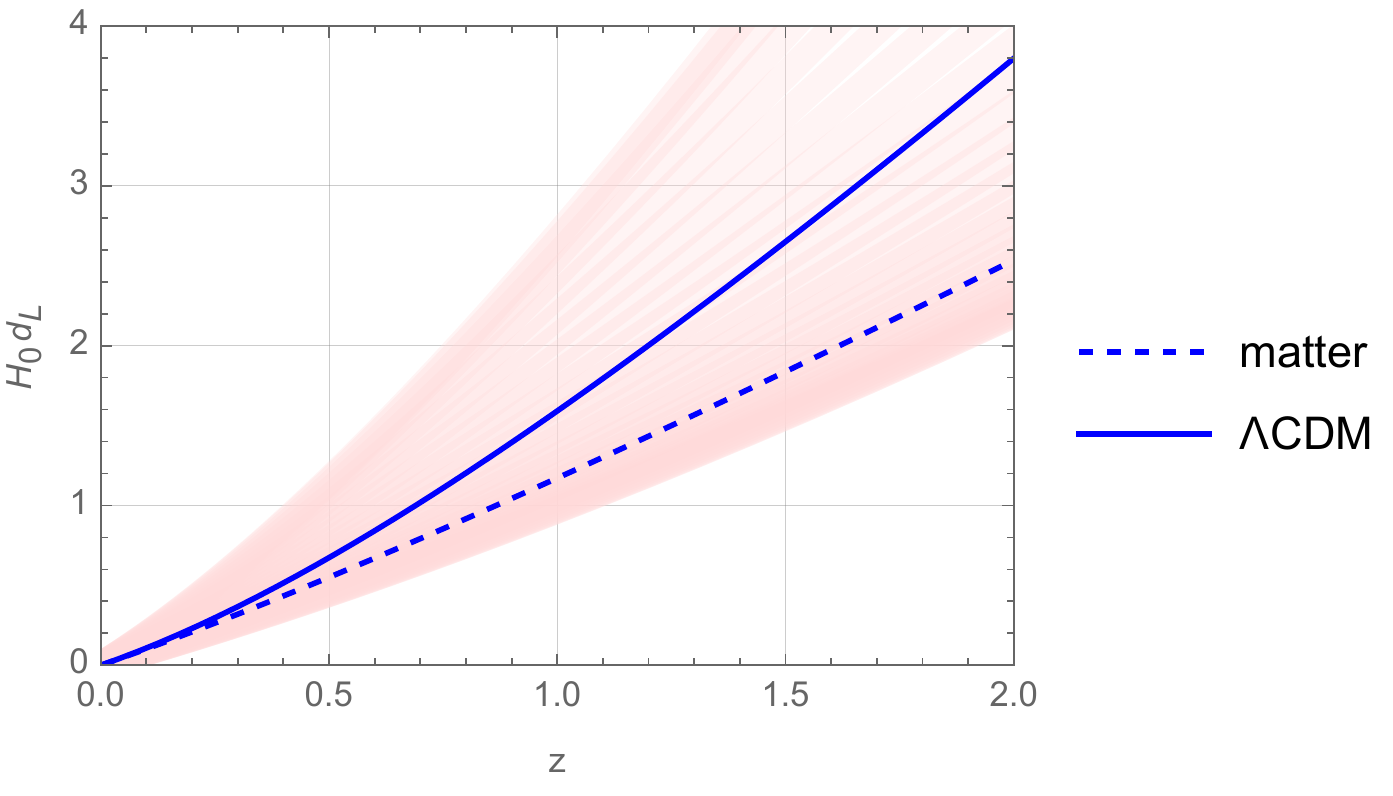}
	\caption{Parametric plot of luminosity distance $H_0d_L$ as a function of $z$ for the open KS model with $\eta_0=3.5$.  The pink band indicates the variation with $\alpha_0$. For comparison, the plot also contains the result for a flat FLRW universe with matter (blue dashed curve) and for a flat $\Lambda$CDM FLRW universe with 25\% matter and 75\% cosmological constant (blue curve). The lensing effect of the anisotropic background can account for the apparent cosmological constant.}
	\label{fig:LD3}
\end{figure}

This can be made more pronounced by increasing $\eta_0$ even further to $\eta_0=3.5$, as in \fref{fig:LD3}. In this case, the $\Lambda$CDM luminosity distance curve falls in the centre of the $H_0d_L(z)$ band corresponding to the variation with $\alpha_0$, as shown in figure~\ref{fig:LD3}.
It is, of course, not clear if such a considerable variation with $\alpha_0$ is consistent with supernova data, although this could be studied in detail~\cite{Dhawan:2022lze,Cowell:2022ehf}. However, as we will now argue, the large anisotropy, $\eta_0={\cal O}(1)$, required for a significant change in the luminosity distance redshift curves, is in conflict with CMB data, at least for solutions of type I. We postpone the discussion of type II solutions to a future publication. 

\subsection{CMB constraints}
For  solutions of type I, in which the universe is isotropic at early times and becomes increasingly anisotropic at later times, the physical observables receive quadrupolar corrections compared to the isotropic case. Even though those recent observations that most strongly describe conflicts with the principles of spatial homogeneity or isotropy are dipolar in nature, quadrupolar modifications are interesting in their own regard, see, for example,~\cite{Heinesen:2021azp,Copi:2010na}. As we will see below, any such anisotropies are heavily constrained by the near-isotropy of the CMB, despite our models being essentially isotropic in the early universe.

It can be argued that the CMB frequency spectrum $n_1(\nu_1)$ is a blackbody spectrum everywhere at the time of decoupling $t_1$
\begin{equation}
	n_1(\nu_1) d\nu_1 = \frac{8\pi \nu_1^2 d\nu_1}{\exp(h\nu_1/(k_B T_1))-1},
\end{equation}
where anisotropies from primordial sources will be $\mathcal O (10^{-5})$ effects. Measuring this today, one needs to account for two effects: the redshift $\nu_1 = (1+z)\nu_0$ of the frequency and an overall volume factor $1/v$ from the expansion of the universe reducing the intensity. Putting this all together, one finds
\begin{eqnarray}
	\eqalign{
	n_0(\nu_0) d\nu_0 &= \frac{n_1((1+z)\nu_0)\, d((1+z)\nu_0)}{v}\\
	&= \frac{(1+z)^3}{v} \frac{8\pi\nu_0^2 d\nu_0}{\exp{(h\nu_0/(k_BT))}-1},
	}
\end{eqnarray}
where $T=T_1/(1+z)$ and
\begin{equation}
	v=\frac{a_0^2b_0}{a_1^2 b_1} = \frac{a_0^3}{a_1^3}\frac{f_1}{f_0}~.
\end{equation}

For the open and closed KS universes with $k=-1$ and, respectively, $k=1$, this spectrum can be written as
\begin{equation}
	n_0(\nu_0) d\nu_0 = (1+F_n) \frac{8\pi\nu_0^2 d\nu_0}{\exp{(h\nu_0/(k_BT))}-1},
\end{equation}
where
\begin{equation}
	1+F_n = \frac{f_0}{f_1} (1+F_z)^3,\quad T = \frac{a_1 T_1}{a_0(1+F_z)}.
\end{equation}
The leading order of the series expansion implies
\begin{eqnarray}
	\eqalign{
	1+F_n &\approx 1  + \frac{k\eta_0^2}{10}(3\sin^2(\alpha)-1),\\
	T &\approx \frac{a_1 T_1}{a_0} \left( 1-\frac{k\eta_0^2}{10}\sin^2(\alpha) \right).
	}
\end{eqnarray}
These are intensity and temperature quadrupole variations of order $k\eta_0^2/10$. Therefore, to be compatible with measurements of the CMB quadrupole, $\eta^2_0 \sim 10^{-4}$, which strongly limits the anisotropy today.

In addition to a temperature quadrupole, CMB anisotropies of all scales inherit structure via a corresponding quadrupolar modulation~\cite{Adamek:2010sg}, that is, correlations between multipoles $\ell$ and $\ell+2$, as well as changes in polarisation~\cite{Pontzen:2007ii}. In this context it is interesting to point out a nuance in the constraining power of the CMB anisotropy on anisotropic cosmologies. For this effect we resort to a common characterisation of anisotropic models of the Bianchi type, the shear amplitude, $\sigma$, which describes the deformation of a fluid element due to anisotropic expansion. Of particular interest is its ratio with a measure of the average expansion rate measured at present time, $(\sigma/H)$, which, in our notation, is of order $\eta^2/30$, as indicated by Eq.~\eref{eq:shearbyexpansion}. An initially decaying background anisotropy implies a larger shear amplitude at recombination than today, so would produce larger polarisation anisotropies, and hence lead to tighter constraints for given polarisation measurements. However, we consider models with growing background anisotropy, where, conversely, measurements of CMB polarisation are less constraining by a factor of $a_0/a_1$. Naively applying the results of~\cite{2016PhRvL.117m1302S} to our model, we would nevertheless find tighter bounds, $\eta^2_0\lesssim10^{-8}$. We remark, however, that such conclusions demand a more detailed calculation, in which also large-scale adiabatic or iso-curvature perturbations may influence the viability of considered models, e.g.~\cite{Domenech:2022mvt}.

To summarize, despite this model being nearly isotropic in the early universe, the evolution of the spectrum till today strongly constrains the current anisotropy. It appears that this problem would be difficult to avoid in any anisotropic cosmological model, with the potential exception of one that gives rise to a dipole only.

\section{Conclusion}\label{sec:conclusions}

With increased precision measurements of astrophysical observables, detailed tests of the $\Lambda$CDM paradigm have become possible. A growing number of observations point towards tensions, which, if taken seriously, indicate the need to revise the standard cosmological model. A~popular avenue to address these tensions is to introduce additional structure in the $\Lambda$CDM model, such as fields with various equations of state whose dynamics affect both the evolution of the background and the CMB fluctuations. 

In this work we have adopted the perspective of relaxing the requirement of spatial isotropy and explored a number of ways in which the analysis of cosmological observations needs to be revisited in this context. In particular we focused on selected models with a growing anisotropy, specifically the closed Kantowski-Sachs (KS) model and the open axisymmetric Bianchi III model, which we referred to as the open Kantowski-Sachs model. The source of anisotropy in these models is essentially topological and can be traced back to the non-vanishing curvature of the maximally-symmetric two-dimensional subspace. A prominent example of KS topology is $S^2\times S^1$. By studying models with growing anisotropy, late time observables may be used to infer the degree of anisotropy present in the universe.

The closed and the open KS models are characterized by two scale factors and we have studied their evolution in universes with radiation, matter and a cosmological constant. In all cases, the two scale factors evolve differently and this source of anisotropy leads to a lensing effect in the propagation of light. We have derived explicit formulae for computing redshifts, angular diameter distances and luminosity distances in these anisotropic universes.

We have shown that an anisotropic universe, when interpreted as an FLRW cosmology can affect the value of the cosmological constant inferred from the dependence of the luminosity distance of type Ia supernovae on the redshift. This fact can have implications for the Hubble tension (the discrepancy between the local measurements of the Hubble parameter and the CMB value), as well as the dipole tension (the discrepancy between the amplitude of the dipole  present in number counts of quasars and the amplitude of the CMB dipole). We have also shown that, in principle, the accelerated expansion of the universe can be seen as an entirely apparent phenomenon caused by the lensing properties of the background. To accomplish this, the value of our order parameter $\eta_0$ (today's value of the time parameter) measuring anisotropy, must be of order $\eta_0^2={\cal O}(1)$. However, this value turns out to be far larger than the bounds inferred from the CMB temperature field.

Indeed, even though the geometry is only modified at late times, the CMB is sensitive to this anisotropy through the line of sight (in a way similar to how the CMB is sensitive to the cosmological constant, which is a late time effect). Consequently, in these models which produce a quadrupolar anisotropy, a large anisotropy is in conflict with the measurement of the CMB anisotropies and constrains $\eta_0$ to satisfy $\eta_0^2<{\cal O}(10^{-4})$. A possible exception are solutions of type II where the value of $f=a/b$ can be arranged to be similar today and at recombination. In this case, the CMB anisotropy due to redshift can be kept small even for values $\eta_0={\cal O}(1)$, while there are substantial anisotropic redshift effect for supernovae which can mimic the effect of a cosmological constant.  A more detailed study of this scenario is postponed to future work.

\section*{Acknowledgments}
AC's research is supported by a Stephen Hawking Fellowship (EPSRC grant EP/T016280/1). TRH is supported by an STFC studentship. SvH acknowledges funding from the Beecroft Trust, and wishes to thank Linacre College for the award of a Junior Research Fellowship. While preparing  Version~1 of this manuscript we became aware of~\cite{Awwad:2022uoz}, whose results largely agree with our account. In Version~2 we acknowledge fruitful discussions with the authors of~\cite{Awwad:2022uoz}, Yassir Awwad and Tomislav Prokopec, and thank Asta Heinesen for clarifications regarding the computations in Section 3.

The authors would like to thank Prateek Agrawal for ongoing helpful discussions and contributions, and Pedro Ferreira and Subir Sarkar for insightful comments.
\vspace{30pt}

\appendix
\section{Null Geodesics}\label{app:Geodesic}
Given a  null geodesic curve $X^\mu(s)$, its tangent vector $V^\mu(s) = dX^\mu(s)/ds$ must satisfy
\begin{eqnarray}
	\eqalign{
	\frac{dV^\mu}{ds} + \Gamma^\mu_{\nu\rho}V^\nu V\rho&=0~,\\
	g_{\mu\nu}V^\mu V^\nu=0~.
	}
\end{eqnarray}
We write the metric $g_{\mu\nu}$ from~\eref{eq:metric} in the form
\begin{eqnarray}\label{metric}
	\eqalign{
	g&=-n(\tau)^2 d\tau^2+b(\tau)^2 g_1+a(\tau)^2 g_2\;,\\
	n(\tau)&=e^{N(\tau)}\;,\quad a(\tau)=e^{A(\tau)}\;,\quad b(\tau)=e^{B(\tau)}\; ,
	}
\end{eqnarray}
where $g_2$ is the maximally symmetric metric on $X_2$,
\begin{eqnarray}
	g_2=	\left\{\begin{array}{ll}r_2^2(d\theta^2+\theta^2d\phi^2)&k=0\\
        			r_2^2(d\theta^2+\sin^2(\theta) d\phi^2)&k=1\\
        			r_2^2(d\theta^2+\sinh^2(\theta) d\phi^2)&k=-1
        		\end{array}\right.
\end{eqnarray} 
and $g_1$ is the maximally symmetric metric on $X_1$:
\begin{equation}
	g_1=r_1^2 d\chi^2\; .
\end{equation} 

The non-zero components of the $g_2$ connection $\Gamma_2$ are given by
\begin{eqnarray}\label{chg2}
	\Gamma_{2,\phi\phi}^\theta&=\left\{\begin{array}{cl}-\theta&\mbox{for }k=0\\-\cos(\theta)\sin(\theta)&\mbox{for }k=1\\-\cosh(\theta)\sinh(\theta)&\mbox{for }k=-1\end{array}\right.,\;\\[4pt]
	\Gamma_{2,\theta\phi}^\phi&=\left\{\begin{array}{cl}\theta^{-1}&\mbox{for }k=0\\\cot(\theta)&\mbox{for }k=1\\\coth(\theta)&\mbox{for }k=-1\end{array}\right.
\end{eqnarray}
and the connection $\Gamma$ for the 4d metric $g$ is
\begin{eqnarray}\label{chg}
	\begin{array}{rclcrcl}
		\Gamma_{\tau\tau}^\tau&=&N'\;,&\quad&\Gamma_{\tau\chi}^\chi&=&B'\;,\\
		\Gamma_{\tau\alpha}^\beta&=&A'\delta_\alpha^\beta\;&\quad&\Gamma_{\alpha\beta}^\tau&=&e^{2A-2N}\,g_{2,\alpha\beta}A'\;,\\
		\Gamma_{\chi\chi}^\tau&=&r_1^2e^{2B-2N}B' &\qquad& \Gamma_{\alpha\beta}^\gamma&=&\Gamma_{2,\alpha\beta}^\gamma\; .
	\end{array} 
\end{eqnarray}
where  the prime denotes $d/d\tau$.
\vspace{8pt}

We write $(X^\mu (s)) =(\tau(s),\chi(s),x^\alpha(s))$, where $x^1 = \theta$ and $x^2=\phi$. The geodesic equation and the null-curve constraint above become
\begin{eqnarray}
	\eqalign{
	\frac{d V^\tau}{ds}{+}N'{V^\tau}^2{+}r_1^2 e^{2B{-}2N}B'{V^\chi}^2{+} e^{2A{-}2N}A'g_{2,\alpha\beta}V^\alpha V^\beta {=} 0\\
	\frac{dV^\chi}{ds} + 2B'V^\tau V^{\chi} =0~,\\
	\frac{d V^\alpha}{ds} + 2 A' V^\tau V^\alpha + \Gamma^\alpha_{2,\beta\gamma}V^\beta V^\gamma =0~,\\
	-e^{2N}{V^{\tau}}^2 + r_1^2 e^{2B}{V^\chi}^2 + e^{2A}g_{2,\alpha\beta}V^\alpha V^\beta =0.
	}
\end{eqnarray}

These equations can be integrated (once) without explicit knowledge of the metric. We start with the third equation (containing the $\phi$- and the $\theta$-geodesic equations) and consider a geodesic in the direction $\theta$, so we set $V^\phi=0$. Given the connection components~\eref{chg2} this means the $\phi$-geodesic equation is satisfied while the $\theta$-equation becomes
\begin{equation}
	\frac{d(e^{2A}V^\theta)}{ds}=0~\Rightarrow~ (V^\alpha)=(V^\theta,V^\phi)=(v_2\, e^{-2A},0)\; ,
\end{equation}
where $v_2$ is a constant. The $\chi$-equation can be integrated immediately and this leads to
\begin{equation}
	\frac{d(e^{2B}V^\chi)}{ds}=0~\Rightarrow~V^\chi=v_1\, e^{-2B}\; ,
\end{equation}
where $v_1$ is another constant. Inserting this into the constraint fixes $V^\tau$ and the complete solution becomes
\begin{eqnarray}\label{Vsol}
	\eqalign{
	{V^\tau}^2 &=r_1^2 v_1^2\, e^{-2B-2N}+r_2^2v_2^2\, e^{-2A-2N}\;,\\
	V^\chi&=v_1\, e^{-2B}\;,\quad
	V^\theta=v_2\, e^{-2A}\;,\quad
	V^\phi=0\; .
	}
\end{eqnarray}
Formally, the solution~\eref{Vsol} can be integrated further for the geodesic solution
\begin{eqnarray}\label{geosol1}
	\eqalign{
	\theta-\theta_0&=v_2\int_{s_0}^{s}ds'\,e^{-2A} =\pm v_2\int_{\tau_0}^\tau d\tau'\,\frac{e^{N-2A}}{\sqrt{r_1^2v_1^2e^{-2B}+r_2^2v_2^2e^{-2A}}}\\
	\chi-\chi_0&=v_1\int_{s_0}^sds'\, e^{-2B} = \pm v_1\int_{\tau_0}^\tau d\tau'\,\frac{e^{N-2B}}{\sqrt{r_1^2v_1^2e^{-2B}+r_2^2v_2^2e^{-2A}}}\; .
	}
\end{eqnarray}

Alternatively, introducing the angle $\alpha$ defined in Section~\ref{sec:Redshift}, these integrals can also be written as
\begin{eqnarray}\label{geosol2}
   \eqalign{
   \theta-\theta_0&=\pm \frac{a_0\cos(\alpha)}{r_2}\int_{\tau_0}^\tau d\tau\, \frac{n(\tau)}{a(\tau)^2(1+z(\tau))}\\
   \chi-\chi_0&=\pm\frac{b_0\sin(\alpha)}{r_1}\int_{\tau_0}^\tau d\tau\,\frac{n(\tau)}{b(\tau)^2(1+z(\tau))}
   }
\end{eqnarray}
where $a_0$ and $b_0$ are the values of the scale factors at $\tau_0$ (`today's time') and $z(\tau)$ is the redshift at time $\tau$, defined in~\eref{redshift}.

The same expressions can also be written as
\begin{eqnarray}
	\eqalign{
	\theta-\theta_0&=\pm \frac{1}{r_2}\int_{\tau_0}^{\tau} d\tau'\,\frac{ n(\tau') /a(\tau')}{\sqrt{1+ \tan (\alpha(\tau'))^2}}\\
	\chi-\chi_0=& \pm \frac{1}{r_1}\int_{\tau_0}^{\tau} d\tau'\,\frac{\tan (\alpha(\tau'))\, n(\tau') /b(\tau')}{\sqrt{1+ \tan (\alpha(\tau'))^2}}~,
	}
\end{eqnarray}
where
\begin{equation}\label{alphadef3}
	\tan(\alpha(\tau))=\frac{r_1v_1a(\tau)}{r_2v_2b(\tau)}\; .
\end{equation}
Defining $f(\tau)=a(\tau)/b(\tau)$, it follows that 
\begin{equation}
	\tan(\alpha(\tau)) = \tan(\alpha_0)\frac{f(\tau)}{f_0}~,
\end{equation}
where $f_0=a(\tau_0)/b(\tau_0)$. Consequently, we have
\begin{eqnarray}\label{eq:theta_chi_expr}
	\eqalign{
	\theta(\tau, \alpha_0)  &{=}\theta_0{\pm}\frac{1}{r_2}\int_{\tau_0}^{\tau} d\tau'\,\frac{  n(\tau') /a(\tau')}{\sqrt{1+ \tan (\alpha_0)^2 (f(\tau')/f_0)^2 } }\\
	\chi(\tau, \alpha_0)  &{=}\chi_0{\pm}\frac{1}{r_1}\int_{\tau_0}^{\tau} d\tau'\,\frac{  \tan (\alpha_0) (f(\tau')/f_0){\cdot}n(\tau') /b(\tau')}{\sqrt{1+ \tan (\alpha_0)^2 (f(\tau')/f_0)^2 } }
	}
\end{eqnarray}

%
\section{Angular Diameter Distance}\label{app:AngDis}
In this section we derive a formula for the angular diameter when the arc along which the distance is measured is contained in the $(\chi,\theta)$-plane. A general formula which takes into account the orientation of the arc can be derived following similar steps (see, e.g.~\cite{Awwad:2022uoz} for a detailed discussion).

Consider two light rays which arrive at $(\chi_0,\theta_0)$ at time~$\tau_0$, with different directional parameters $\alpha_0$ and $\tilde \alpha_0=\alpha_0+\delta\alpha_0$.
What angle do these light rays form at $\chi_0,\eta_0$? To work this out, consider the two tangent vectors
\begin{equation}
	V(\tau_0) = \pm n_0
	\left(\begin{array}{@{}*{20}{c}@{}}
	\displaystyle\frac{\sin(\alpha_0)}{r_1b_0} \\[8pt]
	\displaystyle\frac{\cos(\alpha_0)}{r_2a_0}
	\end{array}\right),
	\quad
	\tilde V(\tau_0) = \pm n_0
	\left(\begin{array}{@{}*{20}{c}@{}}
		\displaystyle\frac{\sin(\tilde\alpha_0)}{r_1b_0} \\[8pt]
		\displaystyle\frac{\cos(\tilde\alpha_0)}{r_2a_0}
	\end{array}\right).
\end{equation}
Defining the spatial metric at $\tau_0$ as $\hat g_0$, the physical angle between the two light rays $\epsilon$ is given by
\begin{eqnarray}
	\cos(\epsilon) = \frac{\hat g_0 (V,\tilde V)}{\sqrt{\hat g_0(V,V) \hat g_0(\tilde V,\tilde V) }} &= \cos(\alpha_0 - \tilde \alpha_0), \\
	&\Rightarrow \epsilon = |\delta\alpha_0|.
\end{eqnarray}

To find the angular diameter distance, we need to trace these null geodesics back to times $\tau_1$ and $\tilde\tau_1=\tau_1+\delta\tau_1$ when light was emitted at $(\chi_1, \theta_1)$ and, respectively, $(\tilde\chi_1, \tilde\theta_1)=(\chi_1+\delta\chi_1, \theta_1+\delta\theta_1)$, where
\begin{eqnarray}
\delta\chi(\tau_1,\alpha_0) &= \left.\frac{\partial \chi}{\partial \tau}\right|_{\tau=\tau_1} \!\!\!\!\!\!\!\!\delta \tau_1 + \frac{\partial \chi}{\partial \alpha_0} \delta \alpha_0~,\qquad \\
\delta\theta(\tau_1,\alpha_0) &=  \left.\frac{\partial \theta}{\partial \tau}\right|_{\tau=\tau_1} \!\!\!\!\!\!\!\!\delta \tau_1 + \frac{\partial \theta}{\partial \alpha_0} \delta \alpha_0~,
\end{eqnarray}
such that the arc between these two locations is perpendicular to the direction of sight, i.e.
\begin{equation}
	\hat{g}_0(V(\tau_1),(\delta\chi_1,\delta\theta_1)) = 0~,   
\end{equation}
which implies the relation 
\begin{equation}
	\frac{\delta\tau_1}{\delta\alpha_0} =  -\frac{ \frac{\tan(\alpha_0)}{f_0} \frac{r_1}{r_2}\frac{\partial \chi}{\partial \alpha_0} +  \frac{\partial \theta}{\partial \alpha_0} }{\frac{\tan(\alpha_0)}{f_0} \frac{r_1}{r_2} \frac{\partial \chi}{\partial \tau} +  \frac{\partial \theta}{\partial \tau} }~.
\end{equation} 
However, this ratio is very small, as it can be checked directly from the cosmological solutions discussed in Section~\ref{sec:EE}. As such, the two rays are essentially  emitted at the same time $\tau_1$.

The physical extension of the object $D$ is then given by
\begin{eqnarray}\label{Ddef}
	D &= \sqrt{r_1^2 b(\tau_1)^2 \delta\chi_1^2 + r_2^2 a(\tau_1)^2 \delta\theta_1^2}\\
	&=~\delta\alpha_0\sqrt{r_1^2 b(\tau_1)^2 \left.\frac{\partial\chi}{\partial\alpha_0}\right|_{\tau=\tau_1}\!\!\!\!\!\!\!\!  + r_2^2 a(\tau_1)^2 \left.\frac{\partial\theta}{\partial\alpha_0}\right|_{\tau=\tau_1}\!\!\!\!\!\!\!\!}
\end{eqnarray}
We can calculate $\delta \chi_1$ and $\delta \theta_1$ using the integrals in~\eref{eq:theta_chi_expr} after plugging in the solutions of interest for the scale factors. The angular diameter distance is then obtained as the ratio
\begin{equation}
	d_A = \frac{D}{\delta\alpha_0}
\end{equation}
which is a function of the time $\tau_1$ and, implicitly, a function of the redshift through~\eref{redshift}.
\vspace{2pt}

More concretely, focusing on the $k=\pm 1$ solutions, for which we used the gauge $n=a$ and the time parameter $\eta =\tau/r_2$, one can consider the quantities $\Delta \theta = \theta_1 - \theta_0$ and $\Delta \chi = \chi_1 - \chi_0$, as 
\begin{eqnarray}\label{eqn:DelTheta}
	r_2 \Delta\theta &= r_2 \cos(\alpha_0)\Delta\eta(1+ F(\alpha,\eta_0,\eta_1)),\\
	\label{eqn:DelChi}
	r_1 \Delta\chi &= r_1 \sin(\alpha_0)\Delta\eta(1+ G(\alpha,\eta_0,\eta_1)),
\end{eqnarray}
where $F$ and $G$, which go to zero in the FRLW limit, are given by
\begin{eqnarray}
	F(\alpha,\eta_0,\eta) &= \frac{1}{\Delta \eta} \int_\eta^{\eta_0} d\eta' \frac{a_0}{a(1+z)}\\
		&= \frac{1}{\Delta \eta} \int_\eta^{\eta_0} d\eta' \frac{1}{\sqrt{\cos^2(\alpha_0) {+} \sin^2(\alpha_0)f^2/f_0^2}} {-}1,\\
	G(\alpha,\eta_0,\eta) &= \frac{1}{\Delta \eta} \int_\eta^{\eta_0} d\eta' \frac{a b_0}{b^2(1+z)}\\
		&= \frac{1}{\Delta \eta} \int_\eta^{\eta_0} d\eta' \frac{f^2/f_0}{\sqrt{\cos^2(\alpha_0) {+} \sin^2(\alpha_0)f^2/f_0^2}} {-}1,
\end{eqnarray}
and $\Delta \eta = \eta_0 - \eta$, $f =a/b$, $f_0 = a_0/b_0$.

The coordinate separation between the two geodesics is given by
\begin{eqnarray}
	r_2 \delta\theta_1 &= r_2(\Delta\theta(\tilde\alpha_0) -\Delta\theta(\alpha_0) ) \\
	& =r_2 \sin(\alpha_0)\epsilon\Delta\eta(1+F(\alpha_0,\eta_0,\eta_1) \\
	&\quad\quad- \cot(\alpha_0)F'(\alpha_0,\eta_0,\eta_1)) + \mathcal O(\epsilon^2),\\[8pt]
	r_1 \delta\chi_1 &= r_1(\Delta\chi(\tilde\alpha_0) -\Delta\chi(\alpha_0) ) \\
	&= r_2 \cos(\alpha_0)\epsilon\Delta\eta(1+G(\alpha_0,\eta_0,\eta_1) \\
	&\quad\quad+ \tan(\alpha_0)G'(\alpha_0,\eta_0,\eta_1)) + \mathcal O(\epsilon^2),
\end{eqnarray}
where the primes indicate a derivative with respect to $\alpha$.

Inserting these results into~\eref{Ddef} gives for the angular diameter distance
\begin{equation}\label{eqn:ADiamDisApp}
	d_A = a_1 r_2 (1+F_A),
\end{equation}
where
\begin{eqnarray}
	1+F_A = &[ \cos^2(\alpha_0) f_1^{-2}(1+G+\tan(\alpha_0)G')^2\\
	&+ \sin^2(\alpha_0)(1+F - \cot(\alpha_0)F')^2]^{1/2}.
\end{eqnarray}

To lowest order in the $\eta$ expansion we should set $F=G=F'=G'=0$ as well as $f_1=1$ so that the entire expression in the square bracket becomes one. This means the lowest order expression for the angular diameter distance is
\begin{equation}
	d_A^{(0)}=a_1r_2\Delta\eta\; .
\end{equation}
In this lowest order limit, where $a=b$, and in the gauge $n=a$, the metric is given by
\begin{equation}\label{localg}
	g=a^2\left(-r_2^2d\eta^2+r_1^2d\chi^2+r_2^2d\theta^2+\cdots\right)
\end{equation}
so $r_2\Delta\eta$ for a light-like geodesic has the simple interpretation of the Euclidean coordinate distance $\Delta R:=\sqrt{r_1^2\Delta\chi^2+r_2^2\Delta\theta^2}$ travelled by the geodesic. Hence, the lowest order angular diameter distance can also be written as $d_A^{(0)}=a_1\Delta R$. 

We can re-write this in terms of the Hubble rate $H_0$, associated to $a$ and defined by
\begin{equation}
	H_0=\frac{1}{r_2a_0^2}\frac{da}{d\eta}(\eta_0)\; .
\end{equation}
This leads to
\begin{equation} \label{H0dA}
	H_0 d_A=\frac{a_1}{a_0^2}\frac{da}{d\eta}(\eta_0)\,\eta_0 \left(1-\frac{\eta_1}{\eta_0}\right) (1+F_A)\; .
\end{equation}
A calculation of $F_A$ to first order in the $\eta_0$ expansion gives
\begin{equation}
	1+F_A= 1+\frac{k}{30}\left(\Delta\eta^2+3\eta_0(\eta_0+\eta_1)\sin^2(\alpha_0)\right)\; ,
\end{equation}
so we have, to this order, 
\begin{eqnarray}
\fl	H_0 d_A=\frac{a_1}{a_0^2}& \frac{da}{d\eta}(\eta_0)\,\eta_0 \left(1-\frac{\eta_1}{\eta_0}\right)\left[1+\frac{k}{30}\left(\Delta\eta^2+3\eta_0(\eta_0+\eta_1)\sin^2(\alpha_0)\right)\right]\; .
\end{eqnarray}


\section*{\refname}

\bibliographystyle{iopart-num}
\bibliography{bibliography}

\providecommand{\newblock}{}
\begin{thebibliography}{10}
\expandafter\ifx\csname url\endcsname\relax
  \def\url#1{{\tt #1}}\fi
\expandafter\ifx\csname urlprefix\endcsname\relax\def\urlprefix{URL }\fi
\providecommand{\eprint}[2][]{\url{#2}}

\bibitem{deSitter:1934}
deSitter W 1934 {\em Proc. Royal Acad. Amsterdam [KNAW]\/} {\bf I} 597--601
  \urlprefix\url{https://www.dwc.knaw.nl/DL/publications/PU00016613.pdf}

\bibitem{1937RSPSA.158..324M}
{Milne} E~A 1937 {\em Proc. R. Soc. Lond. Ser. A\/} {\bf 158} 324--348

\bibitem{Planck:2018vyg}
Aghanim N {\em et~al.\/} (Planck) 2020 {\em Astron. Astrophys.\/} {\bf 641} A6
  [Erratum: Astron.Astrophys. 652, C4 (2021)] (\textit{Preprint}
  \eprint{1807.06209})

\bibitem{Bertone:2004pz}
Bertone G, Hooper D and Silk J 2005 {\em Phys. Rept.\/} {\bf 405} 279--390
  (\textit{Preprint} \eprint{hep-ph/0404175})

\bibitem{Copeland:2006wr}
Copeland E~J, Sami M and Tsujikawa S 2006 {\em Int. J. Mod. Phys. D\/} {\bf 15}
  1753--1936 (\textit{Preprint} \eprint{hep-th/0603057})

\bibitem{Peebles:2002gy}
Peebles P~J~E and Ratra B 2003 {\em Rev. Mod. Phys.\/} {\bf 75} 559--606
  (\textit{Preprint} \eprint{astro-ph/0207347})

\bibitem{Sarkar:2022qgb}
Sarkar S 2022 {\em Inference: International Review of Science\/}
  \urlprefix\url{https://inference-review.com/article/heart-of-darkness}

\bibitem{Perivolaropoulos:2021jda}
Perivolaropoulos L and Skara F 2022 {\em New Astron. Rev.\/} {\bf 95} 101659
  (\textit{Preprint} \eprint{2105.05208})

\bibitem{DiValentino:2021izs}
Di~Valentino E, Mena O, Pan S, Visinelli L, Yang W, Melchiorri A, Mota D~F,
  Riess A~G and Silk J 2021 {\em Class. Quant. Grav.\/} {\bf 38} 153001
  (\textit{Preprint} \eprint{2103.01183})

\bibitem{Copi:2010na}
Copi C~J, Huterer D, Schwarz D~J and Starkman G~D 2010 {\em Adv. Astron.\/}
  {\bf 2010} 847541 (\textit{Preprint} \eprint{1004.5602})

\bibitem{Abdalla:2022yfr}
Abdalla E {\em et~al.\/} 2022 {\em JHEAp\/} {\bf 34} 49--211 (\textit{Preprint}
  \eprint{2203.06142})

\bibitem{Lopez:2022kbz}
Lopez A~M, Clowes R~G and Williger G~M 2022 {\em Mon. Not. Roy. Astron. Soc.\/}
  {\bf 516} 1557--1572 (\textit{Preprint} \eprint{2201.06875})

\bibitem{Yadav:2010cc}
Yadav J~K, Bagla J~S and Khandai N 2010 {\em Mon. Not. Roy. Astron. Soc.\/}
  {\bf 405} 2009 (\textit{Preprint} \eprint{1001.0617})

\bibitem{Migkas:2020fza}
Migkas K, Schellenberger G, Reiprich T~H, Pacaud F, Ramos-Ceja M~E and Lovisari
  L 2020 {\em Astron. Astrophys.\/} {\bf 636} A15 (\textit{Preprint}
  \eprint{2004.03305})

\bibitem{Migkas:2021zdo}
Migkas K, Pacaud F, Schellenberger G, Erler J, Nguyen-Dang N~T, Reiprich T~H,
  Ramos-Ceja M~E and Lovisari L 2021 {\em Astron. Astrophys.\/} {\bf 649} A151
  (\textit{Preprint} \eprint{2103.13904})

\bibitem{Blake:2002gx}
Blake C and Wall J 2002 {\em Nature\/} {\bf 416} 150--152 (\textit{Preprint}
  \eprint{astro-ph/0203385})

\bibitem{Rubart:2013tx}
Rubart M and Schwarz D~J 2013 {\em Astron. Astrophys.\/} {\bf 555} A117
  (\textit{Preprint} \eprint{1301.5559})

\bibitem{Secrest:2020has}
Secrest N~J, von Hausegger S, Rameez M, Mohayaee R, Sarkar S and Colin J 2021
  {\em Astrophys. J. Lett.\/} {\bf 908} L51 (\textit{Preprint}
  \eprint{2009.14826})

\bibitem{Secrest:2022uvx}
Secrest N~J, von Hausegger S, Rameez M, Mohayaee R and Sarkar S 2022 {\em
  Astrophys. J. Lett.\/} {\bf 937} L31 (\textit{Preprint} \eprint{2206.05624})

\bibitem{Dam:2022wwh}
Dam L, Lewis G~F and Brewer B~J 2023 {\em Mon. Not. Roy. Astron. Soc.\/} {\bf
  525} 231--245 (\textit{Preprint} \eprint{2212.07733})

\bibitem{Schwarz:2007wf}
Schwarz D~J and Weinhorst B 2007 {\em Astron. Astrophys.\/} {\bf 474} 717--729
  (\textit{Preprint} \eprint{0706.0165})

\bibitem{Javanmardi:2015sfa}
Javanmardi B, Porciani C, Kroupa P and Pflamm-Altenburg J 2015 {\em Astrophys.
  J.\/} {\bf 810} 47 (\textit{Preprint} \eprint{1507.07560})

\bibitem{Colin:2019opb}
Colin J, Mohayaee R, Rameez M and Sarkar S 2019 {\em Astron. Astrophys.\/} {\bf
  631} L13 (\textit{Preprint} \eprint{1808.04597})

\bibitem{Aluri:2022hzs}
Aluri P~K {\em et~al.\/} 2023 {\em Class. Quant. Grav.\/} {\bf 40} 094001
  (\textit{Preprint} \eprint{2207.05765})

\bibitem{Misner1968}
{Misner} C~W 1968 {\em \APJ\/} {\bf 151} 431

\bibitem{Collins1973}
{Collins} C~B and {Hawking} S~W 1973 {\em \APJ\/} {\bf 180} 317--334

\bibitem{Bolejko_2011}
Bolejko K, C\'el\'erier M~N and Krasi\'nski A 2011 {\em Classical and Quantum
  Gravity\/} {\bf 28} 164002
  \urlprefix\url{https://dx.doi.org/10.1088/0264-9381/28/16/164002}

\bibitem{Pontzen:2010eg}
Pontzen A and Challinor A 2011 {\em Class. Quant. Grav.\/} {\bf 28} 185007
  (\textit{Preprint} \eprint{1009.3935})

\bibitem{Jaffe_2005}
Jaffe T~R, Banday A~J, Eriksen H~K, G\'orski K~M and Hansen F~K 2005 {\em The
  Astrophysical Journal\/} {\bf 629} L1
  \urlprefix\url{https://dx.doi.org/10.1086/444454}

\bibitem{Pontzen:2007ii}
Pontzen A and Challinor A 2007 {\em Mon. Not. Roy. Astron. Soc.\/} {\bf 380}
  1387--1398 (\textit{Preprint} \eprint{0706.2075})

\bibitem{Planck:2015gmu}
Ade P~A~R {\em et~al.\/} (Planck) 2016 {\em Astron. Astrophys.\/} {\bf 594} A18
  (\textit{Preprint} \eprint{1502.01593})

\bibitem{Russell_2014}
Russell E, Kilinc C~B and Pashaev O~K 2014 {\em Monthly Notices of the Royal
  Astronomical Society\/} {\bf 442} 2331--2341 ISSN 0035-8711
  (\textit{Preprint}
  \eprint{https://academic.oup.com/mnras/article-pdf/442/3/2331/3558602/stu932.pdf})
  \urlprefix\url{https://doi.org/10.1093/mnras/stu932}

\bibitem{Akarsu:2019pwn}
Akarsu O, Kumar S, Sharma S and Tedesco L 2019 {\em Phys. Rev. D\/} {\bf 100}
  023532 (\textit{Preprint} \eprint{1905.06949})

\bibitem{Akarsu:2021max}
Akarsu O, Di~Valentino E, Kumar S, Ozyigit M and Sharma S 2023 {\em Phys. Dark
  Univ.\/} {\bf 39} 101162 (\textit{Preprint} \eprint{2112.07807})

\bibitem{2016PhRvL.117m1302S}
{Saadeh} D, {Feeney} S~M, {Pontzen} A, {Peiris} H~V and {McEwen} J~D 2016 {\em
  \PRL\/} {\bf 117} 131302 (\textit{Preprint} \eprint{1605.07178})

\bibitem{Saadeh:2016bmp}
Saadeh D, Feeney S~M, Pontzen A, Peiris H~V and McEwen J~D 2016 {\em Mon. Not.
  Roy. Astron. Soc.\/} {\bf 462} 1802--1811 (\textit{Preprint}
  \eprint{1604.01024})

\bibitem{Kompanyeets:1964}
Kompanyeets A~S and Chernov A~S 1964 {\em Sov.~Phys.~JETP\/} {\bf 20} 1303

\bibitem{Kantowski:1966te}
Kantowski R and Sachs R~K 1966 {\em J. Math. Phys.\/} {\bf 7} 443

\bibitem{Collins:1977fg}
Collins C~B 1977 {\em J. Math. Phys.\/} {\bf 18} 2116

\bibitem{wainwrightellis1997}
Wainwright J and Ellis G (eds) 1997 {\em {Dynamical Systems in Cosmology}\/}
  (Cambridge University Press)

\bibitem{Ellis:1998ct}
Ellis G~F~R and van Elst H 1999 {\em NATO Sci. Ser. C\/} {\bf 541} 1--116
  (\textit{Preprint} \eprint{gr-qc/9812046})

\bibitem{ellis_maartens_maccallum_2012}
Ellis G~F~R, Maartens R and MacCallum M~A~H 2012 {\em Relativistic Cosmology\/}
  (Cambridge University Press)

\bibitem{Barrow:1996gx}
Barrow J~D and Dabrowski M~P 1997 {\em Phys. Rev. D\/} {\bf 55} 630--638
  (\textit{Preprint} \eprint{hep-th/9608136})

\bibitem{Adamek:2010sg}
Adamek J, Campo D and Niemeyer J~C 2010 {\em Phys. Rev. D\/} {\bf 82} 086006
  (\textit{Preprint} \eprint{1003.3204})

\bibitem{Stephani:2003tm}
Stephani H, Kramer D, MacCallum M~A~H, Hoenselaers C and Herlt E 2003 {\em
  {Exact solutions of Einstein's field equations}\/} Cambridge Monographs on
  Mathematical Physics (Cambridge: Cambridge Univ. Press) ISBN
  978-0-521-46702-5, 978-0-511-05917-9

\bibitem{Bradley:2011rt}
Bradley M, Dunsby P~K~S, Forsberg M and Keresztes Z 2012 {\em Class. Quant.
  Grav.\/} {\bf 29} 095023 (\textit{Preprint} \eprint{1106.4932})

\bibitem{Tomita_68}
Tomita K 1968 {\em Progress of Theoretical Physics\/} {\bf 40} 264--276 ISSN
  0033-068X (\textit{Preprint}
  \eprint{https://academic.oup.com/ptp/article-pdf/40/2/264/5305361/40-2-264.pdf})
  \urlprefix\url{https://doi.org/10.1143/PTP.40.264}

\bibitem{Saunders_69}
Saunders P~T 1969 {\em Monthly Notices of the Royal Astronomical Society\/}
  {\bf 142} 213--227 ISSN 0035-8711 (\textit{Preprint}
  \eprint{https://academic.oup.com/mnras/article-pdf/142/2/213/9360709/mnras142-0213.pdf})
  \urlprefix\url{https://doi.org/10.1093/mnras/142.2.213}

\bibitem{Schucker:2014wca}
Schucker T, Tilquin A and Valent G 2014 {\em Mon. Not. Roy. Astron. Soc.\/}
  {\bf 444} 2820--2836 (\textit{Preprint} \eprint{1405.6523})

\bibitem{Fleury:2014rea}
Fleury P, Pitrou C and Uzan J~P 2015 {\em Phys. Rev. D\/} {\bf 91} 043511
  (\textit{Preprint} \eprint{1410.8473})

\bibitem{Awwad:2022uoz}
Awwad Y and Prokopec T 2022  (\textit{Preprint} \eprint{2211.16893})

\bibitem{Dhawan:2022lze}
Dhawan S, Borderies A, Macpherson H~J and Heinesen A 2023 {\em Mon. Not. Roy.
  Astron. Soc.\/} {\bf 519} 4841--4855 (\textit{Preprint} \eprint{2205.12692})

\bibitem{Cowell:2022ehf}
Cowell J~A, Dhawan S and Macpherson H~J 2023 {\em Monthly Notices of the Royal
  Astronomical Society\/}  stad2788 ISSN 0035-8711 (\textit{Preprint}
  \eprint{2212.13569}) \urlprefix\url{https://doi.org/10.1093/mnras/stad2788}

\bibitem{Heinesen:2021azp}
Heinesen A and Macpherson H~J 2022 {\em JCAP\/} {\bf 03} 057 (\textit{Preprint}
  \eprint{2111.14423})

\bibitem{Domenech:2022mvt}
Dom\`enech G, Mohayaee R, Patil S~P and Sarkar S 2022 {\em JCAP\/} {\bf 10} 019
  (\textit{Preprint} \eprint{2207.01569})

\end{thebibliography}

\end{document}